\documentclass[preprintnumbers,amsmath,amssymb,onecolumn]{revtex4}
\usepackage[driverfallback=dvipdfm]{hyperref}
\usepackage{color}
\usepackage{graphicx}
\usepackage{makeidx}
\usepackage{bm}
\usepackage[title]{appendix} 
\usepackage{tabu}
\usepackage{float}
\usepackage{chemarr}

\begin{document}
	\title{\large Thermodynamic Uncertainty Relation in the interlinked cascade of RabGTPases}
	\author{Athokpam Langlen Chanu$^{1}$ and R.K. Brojen Singh$^{1}$}
	\email{brojen@jnu.ac.in (Corresponding author)}
	\affiliation{ \\ $^{1}$School of Computational And Integrative Sciences, Jawaharlal Nehru University, New Delhi-110067, India.}
	\begin{abstract}
		{\noindent}We model the well-known interlinked cascade of Rab GTPases found in eukaryotic cells by using a network of Markov states to investigate the universal Thermodynamic Uncertainty Relation for the non-equilibrium system. First, we prove numerically the TUR in both single Rab species model and the interlinked two Rab species model. Our results show that when two Rab GTPase proteins are interlinked at far from equilibrium, the thermodynamic cost and hence precision is greatly enhanced as compared to single species switching. This implies that at far from equilibrium, the proteins tries to optimise the precision of the performance of their biomolecular processes by forming interlinks in the cascade. Again,our results imply that the interlinked cascade (or oscillator) can achieve a range of tunable rate constants (or frequencies) which suggests a means of maintaining its robustness. Lastly, we highlight a close relation between thermodynamic cost-precision, triangular motifs and disease dynamics.   
	\end{abstract}
	%PACS number(s): 71.23, 72.15.R, 73.50
	\maketitle
\section{Introduction}
\label{sec:5.1}
\begin{figure}[H]
	\centering
	\includegraphics[height=7cm,width=11cm]{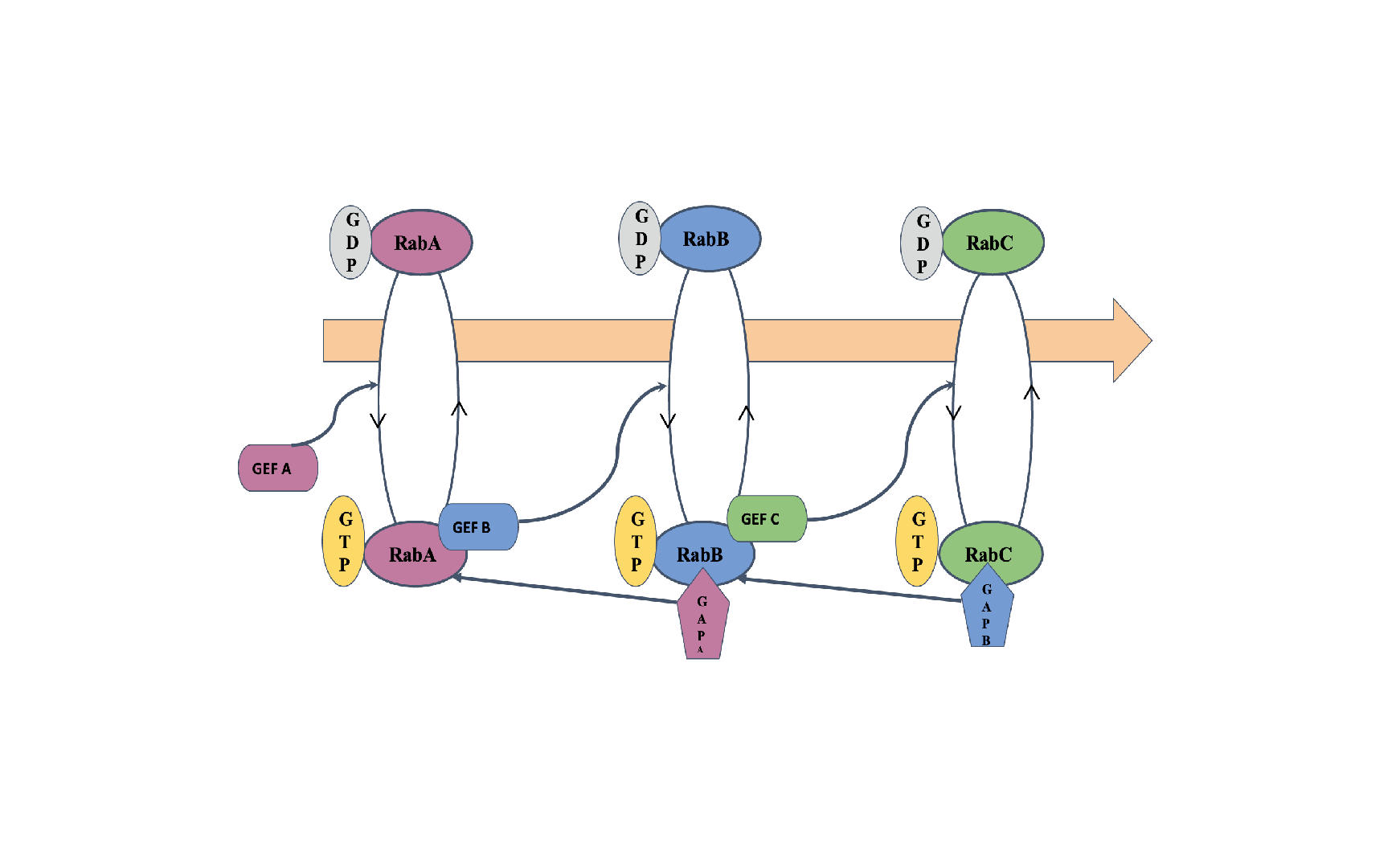}
	\vspace{-1.5cm}
	\caption{\small \textbf{The guanine nucleotide exchange factor (GEF) and GTPase activating protein (GAP) cascades of Rab GTPases:} A GEF specific to the first RabGTPase (RabA) catalyses its activation from the GDP-bound inactive state to the GTP-bound active state. The active RabA then interacts with its co-factor proteins and catalyses the activation of the downstream Rab (RabB). Now, the active GTP-bound RabB has two functions: first, it binds the GAP of the upstream RabA to inactivate this upstream RabA and secondly, it activates the next downstream Rab (RabC) in the cascade.  Figure is adapted from the reference \cite{hutagalung}.}
	\label{fig:mechanism}
\end{figure}

\noindent Complex networks are studied across many disciplines, including information technology, biochemistry, network biology, neuroscience, social networks and ecology. Patterns of interconnections called network motifs are considered to be the basic building blocks of such complex networks \cite{alonscience,alon}. Although the structure of networks in these different disciplines varies, the fundamental network motifs are the same. Thus, network motifs are used to study structural design principles of any complex network \cite{alon}. One such class of network motifs found in cell signalling networks is the interlinked cascades, coupled through positive and negative feedback loops. A well-known example of interlinked cascades is the GTPase cascades found in many parts of eukaryotic cells. Some examples of GTPase cascades include Rab5-HOPS-Rab7, Rab5-SAND-1/Mon1-Rab7 and Rab22-Rabex-5-Rab5 cascades on endosomal traffic, and Ypt1p-Ypt32p, Ypt32p-Sec2p-Sec4p, Rab33b-Rab6 and Rab11-Rabin8-Rab8 cascades on secretory pathways \cite{mizuno}. In this work, we study the Rab GTPases cascade particularly. The Rab is a sub-family member of the Ras (rat sarcoma) superfamily of small GTP (guanosine triphosphate) proteins \cite{mizuno,barr}. Other sub-families include Rho, Arf and Ran \cite{colicelli}. Rab family proteins are involved in regulating signal transduction and in key cellular processes such as cell differentiation, proliferation, cell motility, membrane trafficking, vesicle transport, nuclear assembly, and cytoskeleton formation \cite{subramani,hutagalung}. Their structure, mechanism, and regulation are well-described in the references \cite{goody,cherfils}. Each Rab species has its own specific set of cofactors and effectors. However, all of them follow the same cascade structure. Rab GTPase cascade consists of small GTP binding proteins. The mechanism of Rab GTPase interlinked cascade (See  Figure $\ref{fig:mechanism}$) is explained as follows. At first, the specific cofactor called guanine nucleotide exchange factor (GEF) catalyses the activation of the first GDP-bound inactive state RabA (say) species. Upon activation, the GTP-bound active state RabA captures its specific cofactor, which catalyses the activation of the second GDP-bound inactive RabB (known as the GEF cascade). Again, the second GTP-bound active RabB captures its specific cofactor, which activates the third GDP-bound inactive RabC species. At the same time, RabB binds to its effector, the GTPase activating proteins (GAP) of the first RabA species that inactivates the first system (known as the GAP cascade). The cascades of activation and deactivation of the Rab GTPases continue in the direction of the arrow as shown in Figure $\ref{fig:mechanism}$. Interlinked cascades show rich dynamics such as switching (bistability) or oscillations \cite{jiang}. They are found to improve the switching quality and are robust to input fluctuations \cite{jiang,ehrmann}.\\ 

{\noindent}Non-equilibrium thermodynamics studies open chemical reaction network systems \cite{zhang,es}. Stochastic thermodynamics, in particular, studies fluctuations in small non-equilibrium systems such as living cells which are biochemical systems \cite{sto1, sto2}. Complex structures such as dissipative structures are achieved at far from equilibrium \cite{prigogine,nicolis}, where we can find interesting regimes such as bistability, excitability and oscillations. The RabGTPase cascades are an open chemical reaction network which shows far from equilibrium dynamics. Biomolecular processes, including those of RabGTPase proteins, generally function at far from equilibrium and are dissipative. These small GTPases work on the free energy consump­tion out of the GTP hydrolysis cycle, which transforms a GTP into a GDP and an inorganic phosphate Pi \cite{ehrmann}. By maintaining an excess of GTP, they work under open, non-equilibrium conditions. The non-zero chemical potential difference (or affinity) is the thermodynamic force that drives the underlying chemical reactions to far from equilibrium. This naturally leads to a free-energy cost. There exists a fundamental relation between the free-energy cost of maintaining such biomolecular processes and the relative uncertainty in the random variable quantifying the output of such processes, given by the recently discovered Thermodynamic Uncertainty Relation (TUR) \cite{barato,gingrich,horowitz,hasegawa}. The TUR states that \textit{for any process running for a time $t$, the product ($Q$) of the total dissipation $(\sigma T t)$ and the square of the relative uncertainty $( \epsilon^2)$ of a generic observable is independent of $t$ and is  bounded by $2k_BT$ i.e., $Q=\sigma Tt\epsilon^2\geq 2k_BT$}.  It requires at least $2\times 10^4$ of free energy to get an uncertainty of one per cent \cite{barato}. The crucial product $Q$ hence provides a trade-off between precision and dissipation. TUR shows that a more precise output requires a higher thermodynamic cost independent of the time used to produce the output. TUR has been studied in various processes such as that of molecular motors \cite{pietzonka,kolomeisky,bustamante}, biochemical oscillations \cite{cao, marsland}, enzymatic cycles \cite{wierenga}, brownian clocks \cite{brownian},kinetic-proofreading \cite{bennett,copy}, sensory adaptation \cite{lan}, glycolytic oscillations\cite{glycolytic,marsland}, regulatory circuits \cite{walczak}, interacting oscillators \cite{lee}.  A detailed TUR to access biological processes can be seen in the references \cite{assess,zhangg}. For a general Markov process with $N$ Markov states, the TUR reads as \cite{barato},
\begin{equation}
	Q\geq \frac{\mathcal{A}}{N} \coth \left(\frac{\mathcal{A}}{2N}\right) \geq 2, \label{eq:q1}
\end{equation}
where $\mathcal{A}$ represents affinity which is the thermodynamic force driving the Markov process.  As $Q$ is an increasing function of $\mathcal{A}$,  the minimum cost for a given uncertainty is obtained when $\mathcal{A}\rightarrow  0$ (i.e, equilibrium). Hence, for $A\rightarrow 0$, $Q$ is minimal  where $Q \rightarrow 2$. The TUR is shown to be valid for general networks, both unicyclic as well as multicyclic networks \cite{barato}. \\

%In equilibrium, i.e., for k+ = k~, there is no dissipation and the uncertainty e diverges. fano factor as a diagnostoc tool for network topology. 
{\noindent}There exists a wealth of literature on the biochemical and structural properties of small GTPase proteins. The dynamical properties of GTPases, such as bistability and oscillations, are studied using response functions such as Michaelis-Menten or Hill functions \cite{ehrmann}. However, very little is known about the non-equilibrium thermodynamics of the interlinked GTPase cascades. Since the interlinked Rab cascade works at far from equilibrium, it is natural to inquire about the cost-precision trade-off in such a non-equilibrium system. In this present chapter, we analyse the single Rab species switching model and a double Rab species interlinked model if the non-equilibrium models obey the TUR. Using analytical and numerical approaches, we prove the TUR for both models. We further investigate how the interlinked system optimises the thermodynamic cost and precision. We then make a connection between the TUR, network motifs and disease dynamics. \\

\noindent The present paper is organized as follows. Section \ref{sec:5.2} presents the theoretical models. The Rab single species switching model is described in section \ref{sec:5.2.1} and the interlinked Rab double species model is presented in section \ref{sec:5.2.2}. Section \ref{sec:5.3} presents the methodology used for analysing the TUR. Sections \ref{sec:5.4.1} and \ref{sec:5.4.2} present the results and discussion. Section \ref{sec:5.5} presents the conclusion. 
\section{Model}
\label{sec:5.2}
\noindent Suppose the interlinked RabGTPases cascade in Figure $\ref{fig:mechanism}$ follows a Markov process. We represent the interlinking mechanism with a network of Markov states (See Figure $\ref{fig:representation}$), where the variables $X_{i}$ and $X_{i}^{*}$ ($i=1,2,\dots , n$) respectively denote the inactive GDP bound-states and the active GTP-bound states of the different Rab proteins. In Figure $\ref{fig:representation}$, the transitions from the inactive to active states occur through positive feedbacks (+ sign in red colour) with their respective GEF cofactors (GEF cascades), and the deactivation of the upstream active states happens via negative feedbacks (- sign in red colour) with their corresponding GAP effectors (GAP cascades).
%Interlinking positive and negative feedback loops can enable proper modulation of signal responses and are robust to input fluctuations \cite{tian, kim1}.  \\
%Interlinking positive and negative feedback loops creates a tunable motif in gene regulatory networks. (tian)\\ \\
%positive negative: we discovered that coupled positive feedbacks enhance signal amplification and bistable characteristics; coupled negative feedbacks realize enhanced homeostasis; coupled positive and negative feedbacks enable reliable decision-making by properly modulating signal responses and effectively dealing with noise. (Kim)\\
\subsection{ Single Rab species switches}
\label{sec:5.2.1}
{\noindent}First, we consider only one species of the Rab protein, say $X_1$ in some volume $V$. That is, in Figure \ref{fig:model}, we consider only the species $X_1$ on the left side (without the interlinking with $X_2$). We now take the simple thermodynamically consistent switching model of a single Rab species given by Ehrmann \textit{et al} \cite{ehrmann}. With the GEF catalyst, the Rab transitions from its inactive GDP-bound state $X_1$ to its active GTP-bound state $X_{1}^{*}$ (GEF cascade). This transition consumes a GTP molecule and releases a GDP molecule. Now, with the action of the GAP effector, the active $X_{1}^{*}$ deactivates to $X_1$ with the release of an inorganic phosphate $P_i$. The Rab species $X_1$ thus acts as a biological switch. In one complete cycle, a GTP molecule is converted into a GDP molecule and an inorganic phosphate $P_i$. The chemical potential difference, $\Delta \mu $, which drives the system to non-equilibrium, is given by,
\begin{equation}
	\Delta \mu =\mu_{GTP}-\mu_{GDP}-\mu_{P_i}.
\end{equation}
The thermodynamic force is the affinity $\mathcal{A}$ which is equal to the chemical potential difference, i.e. $\mathcal{A}=\Delta \mu$. The reactions involved in the thermodynamically consistent switching model are as follows \cite{ehrmann}. 
\begin{align}
	X_{1}^{*} &\underset{w_{1}^{-}}{\stackrel{w_{1}^{+}}{\rightleftharpoons}} X_1+P_i \ \ \ \ \ \   \ \ \ \ \ \  \ \  \ \ \ \ (Hydrolysis \ Reaction) \label{eq:hr}\\
	I_1+2X^{*}_{1}+X_1+GTP&\underset{k_{11}^{-}}{\stackrel{k_{11}^{+}}{\rightleftharpoons}} I_1+3X_{1}^{*}+GDP  \ \ \ \ \  \ (Nucleotide \ Exchange  \ Reaction) 
	\label{eq:ner} 
\end{align}
The input $I_1$ represents the catalyst GEF molecule specific to the Rab species $X_1$. Suppose all the rate constants in the reactions \eqref{eq:hr} and \eqref{eq:ner} of the switching model are in $time^{-1}$ units. The rate constant $k_{11}^{+}$ represents a self-feedback or activation. By the local-detailed balance condition, we write,
\begin{equation}
	\left(\frac{\Gamma_{+}}{\Gamma_{-}}\right)= \left(\frac{k_{11}^{+}w_{1}^{+}}{k_{11}^{-}w_{1}^{-}}\right) =\exp\left (\frac{\Delta \mu }{k_BT}\right)=e^{\Delta \mu}=e^{\mathcal{A}},
\end{equation}
where $k_B$ and $T$ represent the Boltzmann’s constant and temperature respectively. $\Gamma_{+}$ and $\Gamma_{-}$ represent the products of forward and backward reaction rate constants respectively. Henceforth, we take $k_B = T = 1$ i.e., dimensionless entropy and energy. 
\subsection{Interlinked cascade between two Rab species}
\label{sec:5.2.2}
\begin{figure}[H]
	\centering
		%\vspace{-3cm}
	\includegraphics[height=7cm,width=11cm,angle=0]{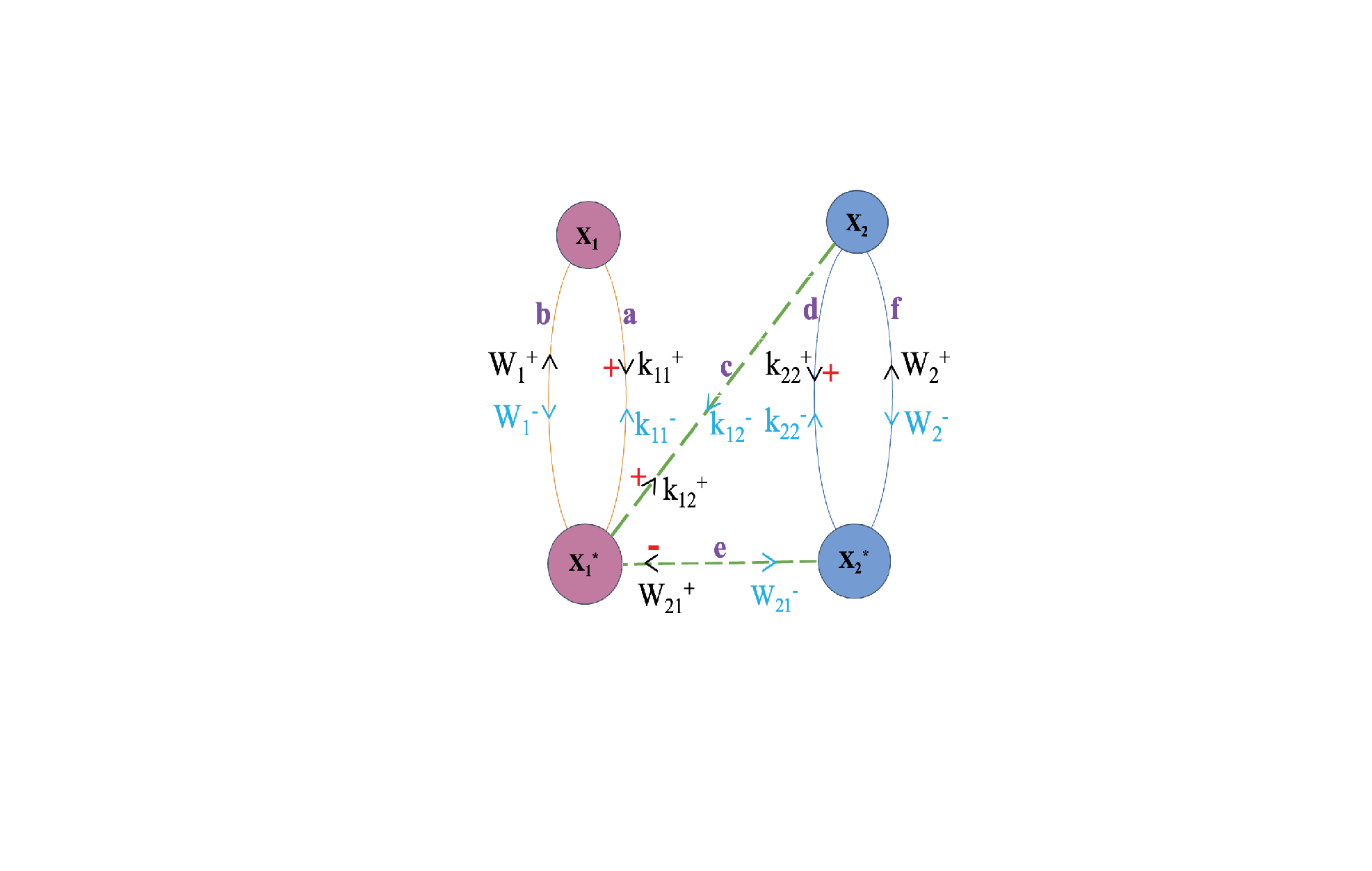}
	\vspace{-1.5cm}
	\caption{\small \textbf{Interlinked model of two Rab species:} $X_1$ and $X_2$ represent two different RabGTPase proteins in their GDP-bound inactive states. The $X_1$ and $X_2$ switch to their respective active states and inactive states following GEF and GAP cascades as explained in Figure \ref{fig:mechanism}.  The green dashed lines indicate the interlinking between $X_1$ and $X_2$ Rab species. Here again, the $+$ signs in red colour indicate a positive feedback and the $-$ signs in red colour indicate a negative feedback.  The letters a,b,c,d,e,f denote the various links in the interconnected network. Rate constants and the directions (arrows) of the reactions are written along the links. For the model system to be thermodynamically consistent (having finite affinity),  all the reaction rates should have their reverse rates in the opposite directions.}
	\label{fig:model}
\end{figure}

{\noindent}In Figure \ref{fig:model}, we introduce a positive feedback (with the rate constant $k_{12}^{+}$) from the active $X_1^{*}$ to the downstream inactive $X_2$ species. Then the inactive $X_2$ activates to $X_2^{*}$ (self-feedfack) with the rate constant $k_{22}^{+}$. Now we introduce a negative feedback from $X_2^{*}$ to $X_1^{*}$ with the rate constant $w_{21}^{+}$. Then $X_1^{*}$ inactivates to $X_1$ with the rate constant $w_1^{+}$. Thermodynamic consistency requires the reversibility of reactions \cite{ehrmann}. In Figure \ref{fig:model}, the symbols $+$ and $-$ in red colours represent the positive and negative feedbacks respectively.  We modify the RabGTPase cascade reactions given by Ehrmann \textit{et al} \cite{ehrmann} in the following way. 
\begin{align}
	I_1+2X_1^{*}+X_1+GTP&\underset{k_{11}^{-}}{\stackrel{k_{11}^{+}}{\rightleftharpoons}} I_1+3X_1^{*}+GDP  \label{eq:one} \\
	X_1^{*} &\underset{w_{1}^{-}}{\stackrel{w_{1}^{+}}{\rightleftharpoons}} X_1+P_i \label{eq:two}\\
	G_{2}^{+}+2X_1^{*}+X_2+GTP&\underset{k_{12}^{-}}{\stackrel{k_{12}^{+}}{\rightleftharpoons}} G_{2}^{+}+3X_2+GDP   \label{eq:three}\\
	G_{2}^{+}+2X_2^{*}+X_2+GTP&\underset{k_{22}^{-}}{\stackrel{k_{22}^{+}}{\rightleftharpoons}} G_{2}^{+}+3X_2^{*}+GDP  \label{eq:four} \\
	X_2^{*} &\underset{w_{2}^{-}}{\stackrel{w_{2}^{+}}{\rightleftharpoons}} X_2+P_i \label{eq:five}\\
	G_{1}^{-}+2X_2^{*}+X_1^{*}&\underset{w_{21}^{-}}{\stackrel{w_{21}^{+}}{\rightleftharpoons}} G_{1}^{-}+2X_2^{*}+P_i \label{eq:six}
\end{align}
Suppose $X_1$ is an RabA and $X_2$ is an RabB. Then $G_2^{+}$ is GEF B and $G_1^{-}$ is GAP A (See Figure \ref{fig:mechanism}). The reactions \eqref{eq:one}, \eqref{eq:three} and \eqref{eq:four} have positive feedbacks, and \eqref{eq:six} has negative feedback. In the reference \cite{ehrmann}, the concept of interlinking is qualitatively described in the reaction channels without an actual interlinking network structure. In the present work, we model the interlinked cascade with a proper network structure of Markov states (See Figures \ref{fig:representation} and \ref{fig:model}) by adding the interlinking reactions \eqref{eq:three} and \eqref{eq:six} in the cascade reactions given in \cite{ehrmann}. In other words, we incorporate activation and inhibition along the interlinks $c$ and $e$ in Figure \ref{fig:model}. The local-detailed balance requires,
\begin{eqnarray}
	\left(\frac{k_{11}^{+}w_{1}^{+}}{k_{11}^{-}w_{1}^{-}}\right) = e^{\mathcal{A}_1}; \
	\left(\frac{k_{12}^{+}k_{22}^{+}w_{21}^{+}}{k_{12}^{-}k_{22}^{-}w_{21}^{-}}\right) = e^{\mathcal{A}_2} ; \
	\left(\frac{k_{22}^{+}w_{2}^{+}}{k_{22}^{-}w_{2}^{-}}\right) = e^{\mathcal{A}_3}, \nonumber 
\end{eqnarray}
where $\mathcal{A}_1$, $\mathcal{A}_2$ and $\mathcal{A}_3$ are the affinities in the three cycles in the interlinked network structure of Figure \ref{fig:model}.
\section{Methods}
\label{sec:5.3}
\noindent Assume a volume $V$ at a fixed temperature $T=1$. Consider inside the volume a Markov process on a general network of $n$ states with the state vector $\textbf{X}=[X_1,X_2,\dots,X_n]^{T}$ (see Figure \ref{fig:representation}). For the single Rab species switching model, $\textbf{X}=[X_1,X_1^{*}]^{T}$, and for the double Rab species interlinked model, $\textbf{X}=[X_1,X_1^{*},X_2,X_2^{*}]^{T}$ (See Figure \ref{fig:model}). The transition rates among the four states of the double Rab species interlinked model of Figure \ref{fig:model} can be seen from the reactions \eqref{eq:one},\eqref{eq:two},\eqref{eq:three},\eqref{eq:four}, \eqref{eq:five} and \eqref{eq:six}. \\

\noindent Let the random variable $Y_{a}$ be some observable of interest along the link $`a$' of the network structure in Figure \ref{fig:model}, for instance the GTP consumption along $`a$' \cite{barato}.
%the output (ATP consumption) along a link of an enzymatic cycle \cite{barato}. 
The product $Q_a$, in the Thermodynamic Uncertainty Relation, of the total dissipation $(\sigma t)$ and the squared relative uncertainty in $Y_a$ (denoted by $\epsilon_a^2$) is given by \cite{barato}
\begin{equation}
	Q_a\equiv \sigma t \epsilon_a^2=\frac{2D_a \sigma }{J_a^2}, \label{eq:product}
\end{equation}
{\noindent}where $D_a$ is the diffusion constant along the link $`a$'; $J_a$ is the stationary probability current along the link  $`a$', and $\sigma= entropy \ production \ rate =\displaystyle \sum_i J_i \mathcal{A}_i$. The $J_i$ represents the stationary probability current along the link $`i$', and $\mathcal{A}_i$ represents the affinity associated with the link $`i$'. We now use the general method developed by Koza \cite{koza, koza1} to compute the stationary probability current $J$ and the diffusion coefficient $D$ in an arbitrary periodic system. For a general network of total $n$ states with transition rates from state $i$ to another state $j$ denoted by $k_{ij}$,  the $n \times n$ generator matrix $\textbf{L}^a(z)$ associated with the observable $Y_a$ is defined as,
\begin{equation}
	\textbf{L}^a(z) =\begin{cases}
		k_{ij}\ e^{z \ d_{ij}} \ \ \ \ \ \ ; & \text{i $\neq$ j}\\
		- \displaystyle \sum_j k_{ij} \ \ \ \ \ \  ; & \text{i= j}.
	\end{cases} \label{eq:lmatrix}
\end{equation}
{\noindent}$d_{ij}$ is the generalised distance which characterises how much the random variable $Y_a$ changes in the $i \rightarrow j$ transition. It is defined as,
\begin{equation}
	d_{ij} = -d_{ji}= \begin{cases}
		1 \ \ \ \ \ \ \ ; & \text{if a product is generated in i $\rightarrow$ j}\\
		0 \  \ \ \ \ \ \ ; & \text{if no product is generated in i $\rightarrow$ j}.
	\end{cases}
\end{equation}
{\noindent}The characteristic polynomial related to the matrix $\textbf{L}(z)$ is defined as,
\begin{equation}
	p(z,y)=det(y\textbf{I}-\textbf{L}(z))=\sum_{n=0}^n C_{n}(z) \ y^n,
\end{equation}
{\noindent}where  $\textbf{I}$ is the identity matrix and $C_{n}(z)$ are the co-efficients of the characteristic polynomial. The characteristic co-efficients $C_{n}(z)$ are functions of transition rates.\\

\noindent The stationary probability current or velocity associated with $Y_a$ is defined as \cite{barato, koza, koza1},
\begin{equation}
	J_a=-\frac{C_{0}^{'}}{C_1}\bigg |_{z=0}. \label{eq:j1}
\end{equation}
{\noindent}According to Koza \cite{koza,koza1}, the diffusion coefficient is defined as, 
\begin{equation}
	D_a=\frac{C_{0}^{''}-2C_{1}^{'}J_a-2C_2J_a^2}{2C_1}\bigg |_{z=0}. \label{eq:koza}
\end{equation}
{\noindent}However, according to Barato and Seifert \cite{barato}, the diffusion coefficient is defined as,
\begin{equation}
	D_a=\frac{-C_{0}^{''}-2C_{1}^{'}J_a-2C_2J_a^2}{2C_1}\bigg |_{z=0}. \label{eq:bs}
\end{equation}
\\ There is a difference in the formula of diffusion coefficient as given by Koza, and Barato and Seifert. We use the following formula modified from the original one given by Koza (i.e., negative of Koza's formula).
\begin{equation}
	D_a=\frac{-C_{0}^{''}+2C_{1}^{'}J_a+2C_2J_a^2)}{2C_1} \bigg |_{z=0}.\label{eq:diff}
\end{equation}
{\noindent}We have analysed our model of sections \ref{sec:5.2.1} and  \ref{sec:5.2.2} with both the formulas \eqref{eq:koza} and \eqref{eq:bs}. We discuss the analysis and justify the use of equation \eqref{eq:diff} in the results and discussion section of \ref{sec:5.4}.\\ 

{\noindent}Fano factor which measures the fluctuation in $Y_a$ is defined as \cite{barato,fanoo},
\begin{equation}
	F_a=\frac{2D_a}{J_a}. \label{eq:fanod}
\end{equation}
Let $P_i(t)$ be the probability of being in the state $i$ of the network at any time $t$. Then the Master equation reads, 
\begin{equation}
	\frac{d}{dt}\textbf{P}=\textbf{L}\textbf{P}, \label{eq:me}
\end{equation}
where the probability state vector $\textbf{P}=\textbf{P}(t)=[P_1(t),P_2(t),\dots , P_n(t)]^{T}$, and $\textbf{L}$ is the stochastic transition matrix and is related to the generator matrix by $\textbf{L}=\textbf{L}(z=0)$.
\section{Results and Discussion}
\label{sec:5.4}
\noindent Using the methodology and formulas described in the above section \ref{sec:5.3}, we present the analysis of the single Rab species switching model and the double Rab species interlinked cascade model in the following subsections. 
\subsection{Single Rab species switching model}
\label{sec:5.4.1}
{\noindent}For the single Rab species switching model described by the reactions \eqref{eq:hr} and \eqref{eq:ner}, the state vector is $\textbf{X}=[X_1,X_1^{*}]^{T}$ (see Figure \ref{fig:model}). Suppose our observable of interest $Y_a$ is the number of GTP molecules consumed or GDP molecules released in the forward reaction of \eqref{eq:ner} with the rate constant $k_{11}^{+}$ (i.e., the link $`a$' in Figure \ref{fig:model}). The generalised distance $d_{12}^a=-d_{21}^a=1$ and $d_{12}^b=d_{21}^b=0$. Using equation \eqref{eq:lmatrix}, we calculate the $2\times 2$ generator matrix $\textbf{L}^a(z)$ for the single species Rab switching model as,
\begin{align}
	\textbf{L}^a(z)
	=\begin{bmatrix}
		$$L_{X_1\rightarrow X_1}$$ & $$L_{X_1\rightarrow X_1^*}$$ \\
		$$L_{X_1^*\rightarrow X_1}$$ & $$L_{X_1*\rightarrow X_1^*}$$
	\end{bmatrix}
	=\begin{bmatrix}
		$$-(k_{11}^{+}+w_{1}^{-})$$ & $$(k_{11}^{+}e^{z}+w_{1}^{-})$$ \\
		$$(k_{11}^{-}e^{-z}+w_{1}^{+})$$ & $$-(k_{11}^{-}+w_{1}^{+})$$ 
	\end{bmatrix}
.
\end{align}
The characteristic equation for $\textbf{L}^a(z)$ is
\begin{align} 
	&|y\textbf{I}-\textbf{L}^a(z)|=0 \nonumber \\
	&\Rightarrow \begin{vmatrix}
		$$y+(k_{11}^{+}+w_{1}^{-})$$ & $$-(k_{11}^{+}e^{z}+w_{1}^{-})$$ \\
		$$-(k_{11}^{-}e^{-z}+w_{1}^{+})$$ & $$y+(k_{11}^{-}+w_{1}^{+})$$ 
	\end{vmatrix}=0 \nonumber \\
	&\Rightarrow [y+(k_{11}^{+}+w_{1}^{-})][y+(k_{11}^{-}+w_{1}^{+})]-(k_{11}^{+}e^{z}+w_{1}^{-})(k_{11}^{-}e^{-z}+w_{1}^{+})=0\nonumber \\
	&\Rightarrow y^2+y(k_{11}^{-}+w_{1}^{+}+k_{11}^{+}+w_{1}^{-})-(k_{11}^{+}k_{11}^{-}+k_{11}^{+}e^{z}w_{1}^{+}+k_{11}^{-}w_{1}^{-}e^{-z}+w_{1}^{+}w_{1}^{-})=0.
\end{align}
Comparing the coefficients of $y^n$ on both sides, we get the following characteristic co-efficients. %$C_2=1;C_1=(k_{11}^{-}+w_{1}^{+}+k_{11}^{+}+w_{1}^{-});C_{1}^{'}=0;C_0=-(k_{11}^{+}k_{11}^{-}+k_{11}^{+}w_{1}^{+}e^{z}+k_{11}^{-}w_{1}^{-}e^{-z}+w_{1}^{+}w_{1}^{-});C_0^{'}=-(k_{11}^{+}e^{z}w_{1}^{+}-k_{11}^{-}w_{1}^{-}e^{-z}); C_0^{''}=-(k_{11}^{+}e^{z}w_{1}^{+}+k_{11}^{-}w_{1}^{-}e^{-z})$.
\begin{align}
	C_2&=1\nonumber \\
	C_1&=(k_{11}^{-}+w_{1}^{+}+k_{11}^{+}+w_{1}^{-})\nonumber \\
	C_{1}^{'}&=0\nonumber \\
	C_0&=-(k_{11}^{+}k_{11}^{-}+k_{11}^{+}w_{1}^{+}e^{z}+k_{11}^{-}w_{1}^{-}e^{-z}+w_{1}^{+}w_{1}^{-})\nonumber \\
	C_0^{'}&=-(k_{11}^{+}e^{z}w_{1}^{+}-k_{11}^{-}w_{1}^{-}e^{-z})\nonumber \\
	C_0^{''}&=-(k_{11}^{+}e^{z}w_{1}^{+}+k_{11}^{-}w_{1}^{-}e^{-z}).\nonumber 
\end{align}
From equation \eqref{eq:j1}, we calculate the stationary probability current as,
\begin{align}
	J_a =-\frac{C_{0}^{'}}{C_1}\bigg |_{z=0}=\frac{k_{11}^{+}e^{z}w_{1}^{+}-k_{11}^{-}w_{1}^{-}e^{-z}}{k_{11}^{+}+w_{1}^{+}+k_{11}^{-}+w_{1}^{-}} \bigg |_{z=0}=\frac{k_{11}^{+}w_{1}^{+}-k_{11}^{-}w_{1}^{-}}{k_{11}^{+}+w_{1}^{+}+k_{11}^{-}+w_{1}^{-}}. \label{eq:j2}
\end{align}
From equation \eqref{eq:diff}, we calculate the diffusion co-efficient as,
\begin{align}
	D_a &=\frac{-(C_{0}^{''}-2C_{1}^{'}J_a-2C_2{J_{a}^{2}})}{2C_1} \bigg |_{z=0} \nonumber \\
	%&=\frac{-\left[-k_{11}^{+}w_{1}^{+}-k_{11}^{-}w_{1}^{-}-2\left(\frac{k_{11}^{+}w_{1}^{+}-k_{11}^{-}w_{1}^{-}}{k_{11}^{+}+w_{1}^{+}+k_{11}^{-}+w_{1}^{-}}\right)^2\right]}{2(k_{11}^{+}+w_{1}^{+}+k_{11}^{-}+w_{1}^{-})}\nonumber \\
	&=\frac{k_{11}^{+}w_{1}^{+}+k_{11}^{-}w_{1}^{-}+2\left(\frac{k_{11}^{+}w_{1}^{+}-k_{11}^{-}w_{1}^{-}}{k_{11}^{+}+w_{1}^{+}+k_{11}^{-}+w_{1}^{-}}\right)^2}{2(k_{11}^{+}+w_{1}^{+}+k_{11}^{-}+w_{1}^{-})}\nonumber \\
	&=\frac{k_{11}^{+}w_{1}^{+}+k_{11}^{-}w_{1}^{-}}{2(k_{11}^{+}+w_{1}^{+}+k_{11}^{-}+w_{1}^{-})}+\frac{ 2\left (k_{11}^{+}w_{1}^{+}-k_{11}^{-}w_{1}^{-}\right)^2}{2(k_{11}^{+}+w_{1}^{+}+k_{11}^{-}+w_{1}^{-})^3}.
\end{align}
From equation \eqref{eq:fanod}, we get the Fano factor as, 
\begin{align}
	F_a=\frac{2D_a}{J_a} %&=\frac{ \frac{k_{11}^{+}w_{1}^{+}+k_{11}^{-}w_{1}^{-}}{(k_{11}^{+}+w_{1}^{+}+k_{11}^{-}+w_{1}^{-})}+\frac{ 2\left (k_{11}^{+}w_{1}^{+}-k_{11}^{-}w_{1}^{-}\right)^2}{(k_{11}^{+}+w_{1}^{+}+k_{11}^{-}+w_{1}^{-})^3} }{\frac{k_{11}^{+}w_{1}^{+}-k_{11}^{-}w_{1}^{-}}{(k_{11}^{+}+w_{1}^{+}+k_{11}^{-}+w_{1}^{-})}}\nonumber \\
	&=\frac{k_{11}^{+}w_{1}^{+}+k_{11}^{-}w_{1}^{-}}{k_{11}^{+}w_{1}^{+}-k_{11}^{-}w_{1}^{-}}+\frac{2(k_{11}^{+}w_{1}^{+}-k_{11}^{-}w_{1}^{-})}{(k_{11}^{+}+w_{1}^{+}+k_{11}^{-}+w_{1}^{-})^2}. \label{eq:f1}
\end{align}
Consider the affinity driving the first cycle with the links $`a$' and $`b$' is $A_I=\mathcal{A}$. We take $k_{11}^{+}=e^{\mathcal{A}/2}$ and $k_{11}^{-}=w_{1}^{+}=w_1^{-}=1$ \cite{pietzonka}. Equation \eqref{eq:f1} becomes,
\begin{align}
	F_a=\frac{e^{\mathcal{A}/2}+1}{e^{\mathcal{A}/2}-1}+\frac{2(e^{\mathcal{A}/2}-1)}{(e^{\mathcal{A}/2}+3)^2}=\frac{e^{\mathcal{A}/4}+e^{-\mathcal{A}/4}}{e^{\mathcal{A}/4}-e^{-\mathcal{A}/4}}+\frac{2(e^{\mathcal{A}/2}-1)}{(e^{\mathcal{A}/2}+3)^2}=\coth(\mathcal{A}/4)+\frac{2(e^{\mathcal{A}/2}-1)}{(e^{\mathcal{A}/2}+3)^2}.  \label{eq:f2}
\end{align}
For the single species rab switching model with $n=2$ states, the probability state vector $\textbf{P}=[P_1(t),P_2(t)]^{T}$. We calculate the stationary probability vector $\textbf{P}^{st}=[P_1,P_2]^{T}$ from the Master equation of \eqref{eq:me} as,
\begin{align}
	&\frac{d}{dt}\textbf{P}^{st}=\textbf{L}^a(z=0) \ \textbf{P}^{st}=0\nonumber \\
	&\implies \begin{bmatrix}
		$$-(k_{11}^{+}+w_{1}^{-})$$ & $$(k_{11}^{+}+w_{1}^{-})$$ \\
		$$(k_{11}^{-}+w_{1}^{+})$$ & $$-(k_{11}^{-}+w_{1}^{+})$$ 
	\end{bmatrix}
	\begin{bmatrix}
		$$P_1$$\\
		$$P_2$$
	\end{bmatrix}=0 \label{eq:probb}	
\end{align}
From equation \eqref{eq:probb}, we find that $P_1 = P_2$.\\ 

\begin{figure}[H]
	\centering
	\includegraphics[height=5cm,width=8cm,angle=0]{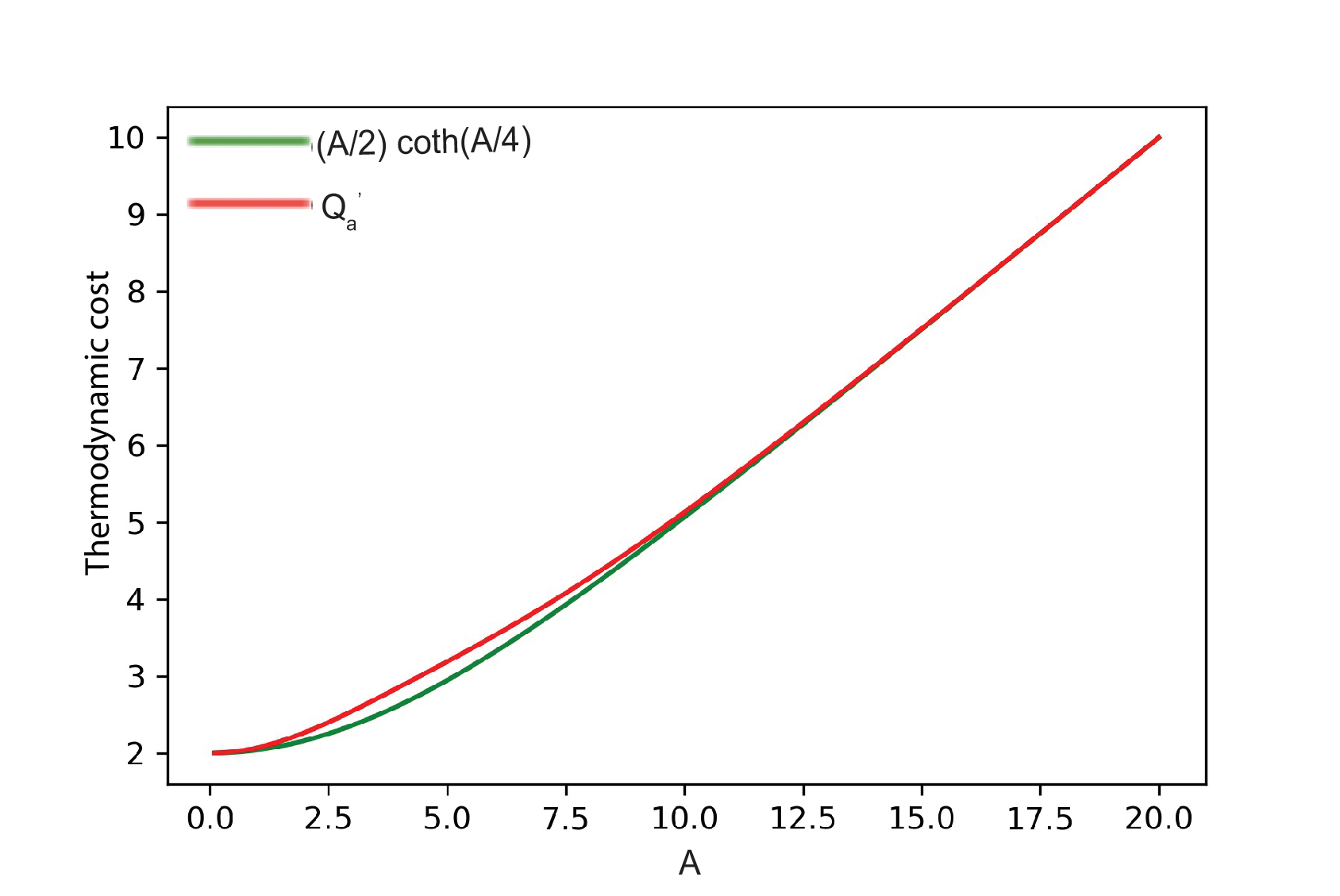}
	\caption{\small \textbf{Numerical result of the single Rab species model}: We see that $Q_a^{'} \rightarrow 2$ as the affinity $\mathcal{A}\rightarrow 0$ and the $Q_a^{'}>2$ as $\mathcal{A}$ increases. This proves the Thermodynamic Uncertainty Relation for the single Rab species model. }
	\label{fig:single result}
\end{figure}

\noindent Now, the entropy production rate in the first cycle is $\sigma=(J_a + J_b)\mathcal{A}.$ The stationary probability current $J_a$ is related to the stationary probabilities by the relation $J_a=\displaystyle \sum_{ij}{d_{ij}^a}(P_ik_{ij}-P_jk_{ji})$. We calculate,
\begin{align}
	J_{a}&=2(P_1k_{11}^{+}-P_2k_{11}^{-}) =2P_1(k_{11}^{+}-k_{11}^{-1})=2P_1(e^{\mathcal{A}/2}-1)\label{eq:jaa}\\
	J_{b}&=2(P_2w_{1}^{+}-P_1w_{1}^{-})=2P_1(w_{1}^{+}-w_{1}^{-})=0\label{eq:jb}
\end{align}
The entropy production rate $\sigma=J_a\mathcal{A}.$ Comparing the expressions of $J_a$ from equations \eqref{eq:j2} and \eqref{eq:jaa}, we get $P_1=\frac{1}{2(e^{\mathcal{A}/2}+3)}$. \\ \\
From equations \eqref{eq:product} and \eqref{eq:f2}, we get,
\begin{equation}
	Q_a= \frac{2D_a\sigma}{J_a^2}=\frac{2D_a}{J_a}\mathcal{A}=F_a\mathcal{A}=\mathcal{A} \coth(\mathcal{A}/4)+\frac{2 \mathcal{A}(e^{\mathcal{A}/2}-1)}{(e^{\mathcal{A}/2}+3)^2}. \label{eq:prob1}
\end{equation}
Dividing the thermodynamic cost $Q_a$ of equation \eqref{eq:prob1} by the total number of states $n=2$ in the model, we get,
\begin{equation}
	Q_a^{'}= \frac{Q_a}{2}=\frac{\mathcal{A}}{2} \coth(\mathcal{A}/4)+\frac{ \mathcal{A}(e^{\mathcal{A}/2}-1)}{(e^{\mathcal{A}/2}+3)^2}.\label{eq:prob}
\end{equation}
The Thermodynamic Uncertainty Relation (TUR) states that the thermodynamic cost $Q \geq \frac{\mathcal{A}}{n} \coth(\frac{\mathcal{A}}{2n}) \geq 2$, where $Q_{min} \rightarrow 2$ as $\mathcal{A} \rightarrow 0$. The minimum cost for a given uncertainty is achieved near equilibrium. We now numerically calculate $Q_a^{'}$ of equation \eqref{eq:prob} for different values of affinity $\mathcal{A}$, as shown in Figure \ref{fig:single result}. From our results, we find that $Q_a^{'}  \rightarrow 2$ as $\mathcal{A} \rightarrow 0$, and $Q_a^{'} > 2$ as $\mathcal{A}$ increases, proving the TUR. For our two-state network model, the TUR of equation \eqref{eq:q1} is $Q_{min}= \frac{\mathcal{A}}{2} \coth(\frac{\mathcal{A}}{4})$. From Figure \ref{fig:single result}, we find that the curve $Q_a^{'}$ of equation \eqref{eq:prob} coincides with the curve of $Q_{min}= \frac{\mathcal{A}}{2} \coth(\mathcal{A}/4)$ as $\mathcal{A} \rightarrow large$. This analysis shows that the Rab species tries to minimise the thermodynamic cost at far away from equilibrium where $\mathcal{A} \neq 0$. However, this situation of minimal thermodynamic cost is not desirable since less thermodynamic cost implies less precision. We now make an interconnected link between the Rab species $X_1$ and another Rab species $X_2$ to investigate the thermodynamic effect of the mechanism of interlinking. 
\subsection{Interlinked cascade model of two Rab species}
\label{sec:5.4.2}
\noindent For the double Rab species model in Figure \ref{fig:model}, the state vector $\textbf{X}=[X_1,X_1^{*},X_2,X_2^{*}]^{T}$. Let us still consider our observable of interest as the output $Y_a$ in the forward reaction of equation \eqref{eq:one} with the rate constant $k_{11}^{+}$ (i.e., along the link $`a$') such that $d_{12}^a=-d_{21}^a=1$. Then, the $4\times 4$ generator matrix $\textbf{L}^a(z)$ along the link $`a$' is given by,
\begin{align}
	&\textbf{L}^a(z)\nonumber \\
	&=\begin{bmatrix}
		$$L_{X_1\rightarrow X_1}$$ & $$L_{X_1\rightarrow X_1^*}$$ & $$L_{X_1\rightarrow X_2}$$ & $$L_{X_1\rightarrow X_2^*}$$\\
		$$L_{X_1^*\rightarrow X_1}$$ & $$L_{X_1^*\rightarrow X_1^*}$$ & $$L_{X_1^*\rightarrow X_2}$$ & $$L_{X_1^*\rightarrow X_2^*}$$\\
		$$L_{X_2\rightarrow X_1}$$ & $$L_{X_2\rightarrow X_1^*}$$ & $$L_{X_2\rightarrow X_2}$$ & $$L_{X_2\rightarrow X_2^*}$$\\
		$$L_{X_2^*\rightarrow X_1}$$ & $$L_{X_2^*\rightarrow X_1^*}$$ & $$L_{X_2^*\rightarrow X_2}$$ & $$L_{X_2^*\rightarrow X_2^*}$$
	\end{bmatrix}
	\nonumber \\
	&=\begin{bmatrix}
		$$-(k_{11}^{+}+w_{1}^{-})$$ & $$(k_{11}^{+}e^{z}+w_{1}^{-})$$ & $0$ & $0$ \\
		$$(k_{11}^{-}e^{-z}+w_{1}^{+})$$ & $$-(k_{11}^{-}+w_{1}^{+}+k_{12}^{+}+w_{21}^{-})$$ & $$k_{12}^{+}$$ &$$w_{21}^{-}$$ \\
		$$0$$ & $$k_{12}^{-}$$ & $$-(k_{12}^{-}+w_{2}^{-}+k_{22}^{+})$$ &$$(k_{22}^{+}+w_{2}^{-})$$ \\
		$$0$$ & $$w_{21}^{+}$$ & $$(k_{22}^{-}+w_{2}^{+})$$ &$$-(k_{22}^{-}+w_{2}^{+}+w_{21}^{+})$$ 
	\end{bmatrix}
.
\end{align}
The characteristic equation of $\textbf{L}^a(z)$ is
\begin{align} 
&|y\textbf{I}-\textbf{L}(z)|=0 \nonumber \\
&\Rightarrow 
\begin{vmatrix}
$$[y+(k_{11}^{+}+w_{1}^{-})]$$ & $$-(k_{11}^{+}e^{z}+w_{1}^{-})$$ & $0$ & $0$ \\
$$-(k_{11}^{-}e^{-z}+w_{1}^{+})$$ & $$[y+(k_{11}^{-}+w_{1}^{+}+k_{12}^{+}+w_{21}^{-})]$$ & $$-k_{12}^{+}$$ &$$-w_{21}^{-}$$ \\
$$0$$ & $$-k_{12}^{-}$$ & $$[y+(k_{12}^{-}+w_{2}^{-}+k_{22}^{+})]$$ &$$-(k_{22}^{+}+w_{2}^{-})$$ \\
$$0$$ & $$-w_{21}^{+}$$ & $$-(k_{22}^{-}+w_{2}^{+})$$ &$$[y+(k_{22}^{-}+w_{2}^{+}+w_{21}^{+})]$$ 
\end{vmatrix} =0. \nonumber \\
\end{align}
%\Rightarrow [y+(k_{11}^{+}+w_{1}^{-})] \begin{vmatrix}
%	$$[y+(k_{11}^{-}+w_{1}^{+}+k_{12}^{+}+w_{21}^{-})]$$ & $$-k_{12}^{+}$$ &$$-w_{21}^{-}$$ \\
%	$$-k_{12}^{-}$$ & $$[y+(k_{12}^{-}+w_{2}^{-}+k_{22}^{+})]$$ &$$-(k_{22}^{+}+w_{2}^{-})$$ \\
%	$$-w_{21}^{+}$$ & $$-(k_{22}^{-}+w_{2}^{+})$$ &$$[y+(k_{22}^{-}+w_{2}^{+}+w_{21}^{+})]$$ 
%	\end{vmatrix}\nonumber \\
%	+(w_1^{+}+k_{11}^{-}e^{-z}) \begin{vmatrix}
%		$$-(k_{11}^{+}e^{z}+w_{1}^{-})$$ & $0$ & $0$ \\
%		$$-k_{12}^{-}$$ & $$[y+(k_{12}^{-}+w_{2}^{-}+k_{22}^{+})]$$ &$$-(k_{22}^{+}+w_{2}^{-})$$ \\
%		$$-w_{21}^{+}$$ & $$-(k_{22}^{-}+w_{2}^{+})$$ &$$[y+(k_{22}^{-}+w_{2}^{+}+w_{21}^{+})]$$ 
%		\end{vmatrix} &=0\nonumber \\
		%&& 
		After calculating the determinants and comparing the co-efficients of $y^n$ on both sides, we calculate, 
		\begin{align}
			C_2&=-(k_{11}^{+}w_1^{+}e^{z}+k_{11}^{-}w_1^{-}e^{-z}+k_{11}^{+}k_{11}^{-}+w_1^{+}w_1^{-})+(k_{12}^{-}k_{22}^{-}+k_{12}^{-}w_{2}^{+}+k_{12}^{-}w_{21}^{+}+k_{22}^{+}w_{21}^{+}\nonumber \\
			&+w_{2}^{-}w_{21}^{+}+k_{11}^{-}k_{12}^{-}+k_{11}^{-}k_{22}^{+}+k_{11}^{-}w_{2}^{-}+k_{11}^{-}k_{22}^{-}+k_{11}^{-}w_{2}^{+}+k_{11}^{-}w_{21}^{+} +w_{1}^{+}k_{12}^{-}+w_{1}^{+}k_{22}^{+}\nonumber \\
			&+w_{1}^{+}w_{2}^{-}+w_{1}^{+}k_{22}^{-}+w_{1}^{+}w_{2}^{+}+w_{1}^{+}w_{21}^{+}+k_{12}^{+}k_{22}^{+}+k_{12}^{+}w_{2}^{-}+k_{12}^{+}k_{22}^{-}+k_{12}^{+}w_{2}^{+}+k_{12}^{+}w_{21}^{+}\nonumber \\
			&+w_{21}^{-}k_{12}^{-}+w_{21}^{-}k_{22}^{+}+w_{21}^{-}w_{2}^{-}+w_{21}^{-}k_{22}^{-}+w_{21}^{-}w_{2}^{+})+(k_{11}^{+}+w_1^{-})(k_{12}^{-}+k_{22}^{+} \nonumber \\
			&+w_2^{-}+k_{22}^{-}+w_2^{+}+w_{21}^{+})+(k_{11}^{+}+w_1^{-})(k_{11}^{-}+w_1^{+}+k_{12}^{+}+w_{21}^{-})\label{eq:c2}
		\end{align}
		\begin{align}
			C_1&=-(k_{11}^{+}w_1^{+}e^{z}+k_{11}^{-}w_1^{-}e^{-z}+k_{11}^{+}k_{11}^{-}+w_1^{+}w_1^{-}) (k_{22}^{-}+w_2^{+}+w_{21}^{+}+k_{12}^{-}+k_{22}^{+}+w_2^{-})\nonumber \\
			& +(k_{11}^{-}+w_1^{+}+k_{12}^{+}+w_{21}^{-})(k_{12}^{-}k_{22}^{-}+k_{12}^{-}w_{2}^{+}+k_{12}^{-}w_{21}^{+}+k_{22}^{+}w_{21}^{+}+w_{2}^{-}w_{21}^{+})\nonumber \\
			& -(k_{12}^{-}k_{12}^{+}k_{22}^{-}+k_{12}^{-}k_{12}^{+}w_{2}^{+}+k_{12}^{-}k_{12}^{+}w_{21}^{+}+k_{12}^{-}w_{21}^{-}k_{22}^{-}+k_{12}^{-}w_{21}^{-}w_{2}^{+}+w_{21}^{+}k_{12}^{+}k_{22}^{+}\nonumber \\
			&+w_{21}^{+}k_{12}^{+}w_{2}^{-}+w_{21}^{+}w_{21}^{-}k_{12}^{-}+w_{21}^{+}w_{21}^{-}k_{22}^{+}+w_{21}^{+}w_{21}^{-}w_{2}^{-}
			) -(k_{11}^{+}+w_1^{-})(k_{12}^{-}k_{12}^{+}+w_{21}^{-}w_{21}^{+}) \nonumber \\
			& +(k_{11}^{+}+w_1^{-})(k_{12}^{-}k_{22}^{-}+k_{12}^{-}w_{2}^{+}+k_{12}^{-}w_{21}^{+}+k_{22}^{+}w_{21}^{+}+w_{2}^{-}w_{21}^{+}) \nonumber \\
			& +(k_{11}^{+}+w_1^{-})(k_{11}^{-}+w_{1}^{+}+k_{12}^{+}+w_{21}^{-})(k_{12}^{-}+k_{22}^{+}+w_{2}^{-}+k_{22}^{-}+w_{2}^{+}+w_{21}^{+}) 
			&\label{eq:c1}
		\end{align}
		\begin{align}
			&C_0=-(k_{11}^{+}w_1^{+}e^{z}+k_{11}^{-}w_1^{-}e^{-z}+k_{11}^{+}k_{11}^{-}+w_1^{+}w_1^{-})(k_{12}^{-}k_{22}^{-}+k_{12}^{-}w_{2}^{+}+k_{12}^{-}w_{21}^{+}+k_{22}^{+}w_{21}^{+}+w_{2}^{-}w_{21}^{+})\nonumber \\
			&+(k_{11}^{+}+w_1^{-})(k_{11}^{-}+w_1^{+}+k_{12}^{+}+w_{21}^{-})
			(k_{12}^{-}k_{22}^{-}+k_{12}^{-}w_{2}^{+}+k_{12}^{-}w_{21}^{+}+k_{22}^{+}w_{21}^{+}+w_{2}^{-}w_{21}^{+} )\nonumber \\
			&-(k_{11}^{+}+w_1^{-})(k_{12}^{+}k_{12}^{-}k_{22}^{-}+k_{12}^{+}k_{12}^{-}w_{2}^{+}+k_{12}^{+}k_{12}^{-}w_{21}^{+}+k_{12}^{-}w_{21}^{-}k_{22}^{-}+k_{12}^{-}w_{21}^{-}w_{2}^{+}\nonumber\\\
			&+w_{21}^{+}k_{12}^{+}k_{22}^{+}+w_{21}^{+}k_{12}^{+}w_{2}^{-}+w_{21}^{+}w_{21}^{-}k_{12}^{-}+w_{21}^{+}w_{21}^{-}k_{22}^{+}+w_{21}^{+}w_{21}^{-}w_{2}^{-}) \label{eq:c0}
		\end{align}
		\begin{align}
			C_{0}^{'}
			%&=(-k_{11}^{+}w_1^{+}e^{z}+k_{11}^{-}w_1^{-}e^{-z})(k_{12}^{-}k_{22}^{-}+k_{12}^{-}w_{2}^{+}+k_{12}^{-}w_{21}^{+}+k_{22}^{+}w_{21}^{+}+w_{2}^{-}w_{21}^{+})\nonumber \\
			&=(-k_{11}^{+}w_1^{+}e^{z}+k_{11}^{-}w_1^{-}e^{-z})K \label{eq:c0prime} \\
			C_{0}^{''}
			%&=(-k_{11}^{+}w_1^{+}e^{z}-k_{11}^{-}w_1^{-}e^{-z})(k_{12}^{-}k_{22}^{-}+k_{12}^{-}w_{2}^{+}+k_{12}^{-}w_{21}^{+}+k_{22}^{+}w_{21}^{+}+w_{2}^{-}w_{21}^{+})\nonumber \\
			&=(-k_{11}^{+}w_1^{+}e^{z}-k_{11}^{-}w_1^{-}e^{-z}) K \label{eq:c0dprime}\\
			C_{1}^{'}
			%&=(-k_{11}^{+}w_1^{+}e^{z}+k_{11}^{-}w_1^{-}e^{-z})(k_{22}^{-}+w_2^{+}+w_{21}^{+}+k_{12}^{-}+k_{22}^{+}+w_2^{-})\nonumber \\
			&=(-k_{11}^{+}w_1^{+}e^{z}+k_{11}^{-}w_1^{-}e^{-z})L \label{eq:c1prime}
		\end{align}
		where $K=(k_{12}^{-}k_{22}^{-}+k_{12}^{-}w_{2}^{+}+k_{12}^{-}w_{21}^{+}+k_{22}^{+}w_{21}^{+}+w_{2}^{-}w_{21}^{+})$ and $L=(k_{22}^{-}+w_2^{+}+w_{21}^{+}+k_{12}^{-}+k_{22}^{+}+w_2^{-})$. \\ \\
		From equation \eqref{eq:j1}, we get the stationary current as,
		\begin{align}
			J_a=-\frac{C_{0}^{'}}{C_1}\bigg |_{z=0}=\frac{(k_{11}^{+}w_1^{+}-k_{11}^{-}w_1^{-})K}{C_1}. \label{eq:j}
		\end{align}
		Using the expressions of $J_a$ and $D_a$ of equations \eqref{eq:j1} and \eqref{eq:diff}, we get the expression of Fano factor as, 
		\begin{align}
			F_a=\frac{2D_a}{J_a}=\frac{C_{0}^{''}}{C_{0}^{'}}- \frac{2C_{1}^{'}J_a}{C_{0}^{'}}- \frac{2J_{a}^{2}C_{2}}{C_{0}^{'}}.\label{eq:ff}
		\end{align}
		{\noindent}Substituting the above expressions, we get,
		\begin{align}
			F_a &=\frac{(k_{11}^{+}w_1^{+}+k_{11}^{-}w_1^{-})}{(k_{11}^{+}w_1^{+}-k_{11}^{-}w_1^{-})} -\frac{2(k_{11}^{+}w_1^{+}-k_{11}^{-}w_1^{-})L}{C_1}+\frac{2(k_{11}^{+}w_1^{+}-k_{11}^{-}w_1^{-})KC_2}{C_{1}^{2}}\nonumber \\
			%&= \frac{(k_{11}^{+}w_1^{+}+k_{11}^{-}w_1^{-})}{(k_{11}^{+}w_1^{+}-k_{11}^{-}w_1^{-})} -\left(\frac{L}{C_1}-\frac{KC_2}{C_{1}^{2}}\right) 2(k_{11}^{+}w_1^{+}-k_{11}^{-}w_1^{-})\nonumber \\
			&=\frac{(k_{11}^{+}w_1^{+}+k_{11}^{-}w_1^{-})}{(k_{11}^{+}w_1^{+}-k_{11}^{-}w_1^{-})} -\left(\frac{LC_1-KC_2}{C_{1}^{2}}\right) 2(k_{11}^{+}w_1^{+}-k_{11}^{-}w_1^{-})\nonumber \\
			&= \frac{e^{\mathcal{A}/4}+1}{e^{\mathcal{A}/4}-1}-\left(\frac{LC_1-KC_2}{C_{1}^{2}}\right) 2(e^{\mathcal{A}/4}-1)\nonumber \\
			&= \frac{e^{\mathcal{A}/8}+e^{-\mathcal{A}/8}}{e^{\mathcal{A}/8}-e^{-\mathcal{A}/8}} -\left(\frac{LC_1-KC_2}{C_{1}^{2}}\right) 2(e^{\mathcal{A}/4}-1)\nonumber \\
			&= \coth{(\mathcal{A}/8)}-\left(\frac{LC_1-KC_2}{C_{1}^{2}}\right) 2(e^{\mathcal{A}/4}-1). \label{eq:coth}
		\end{align}
		The affinity $\mathcal{A}=0$ implies an equilibrium condition.  The total entropy production rate along all the links in the interlinked network of Figure \ref{fig:model} is $\sigma=J_IA_I+J_{II}A_{II}+J_{III}A_{III}$. The $A_{I}$, $A_{II}$ and $A_{III}$ are the affinities driving the cycles with links $(a-b)$, $(c-d-e)$ and $(d-f)$ respectively (see Figure \ref{fig:model}). The stationary probability currents $J_{I},J_{II} $ and $J_{III}$ are non-zero and are given by $J_I=(J_a+J_b)$, $J_{II}=(J_c+J_d+J_e)$ and $J_{III}=(J_d+J_f)$.  \\ 
		
		\noindent Using equation\eqref{eq:coth}, the thermodynamic cost $Q_a$ of equation \eqref{eq:product} becomes,
		\begin{align}
			Q_a= \frac{2D_a \sigma}{J_{a}^{2}}&= \frac{2D_a }{J_{a}^{2}} [(J_a+J_b)A_{I}+(J_c+J_d+J_e)A_{II}+(J_d+J_f)A_{III}] \nonumber \\
			&= \frac{2D_a }{J_{a}}A_I+ \frac{2D_a }{J_{a}^2}[J_bA_{I}+(J_c+J_d+J_e)A_{II}+(J_d+J_f)A_{III}] \nonumber \\
			& =A_{I}\bigg[\coth{(\mathcal{A}/8)} -\left(\frac{LC_1-KC_2}{C_{1}^{2}}\right) 2(e^{\mathcal{A}/4}-1) \bigg]\nonumber \\
			&+ \frac{2D_a }{J_{a}^2}[J_bA_{I}+(J_c+J_d+J_e)A_{II}+(J_d+J_f)A_{III}] \label{eq:finale}
		\end{align}
		By referring to \cite{pietzonka}, we take $k_{11}^{+}=k_{22}^{+}=e^{\mathcal{A}/4}; k_{11}^{-}=k_{22}^{-}=1;w_{1}^{+}=w_{1}^{-}=w_{2}^{+}=w_{2}^{-}=1;k_{12}^{+}=ke^{3\mathcal{A}/4};k_{12}^{-}=\frac{1}{k};w_{21}^{+}=\frac{e^{3\mathcal{A}/4}}{k};w_{21}^{-}=k$. We take these parameters taking into account the ideas such as linking fast and slow positive feedback loops creates an optimal bistable switch in cell signaling \cite{fastandslow}, and the possibility of differences in the reaction rates of activation and repression. The $k$ is an arbitrary constant parameter.\\ 
		
		\noindent As before, we calculate the stationary probability currents along all the links in the interlinked system of Figure \ref{fig:model} as
		% viz., $J_a=\displaystyle \sum_{ij}{d_{ij}^a}(P_ik_{ij}-P_jk_{ji})$. Using this, we calculate,
		$J_{a}=2(P_1k_{11}^{+}-P_2k_{11}^{-} );
		J_{b}=2(P_2w_{1}^{+}-P_1w_{1}^{-});
		J_{c}=2(P_2k_{12}^{+}-P_3k_{12}^{-});
		J_{d}=2(P_3k_{22}^{+}-P_4k_{22}^{-});
		J_{e}=2(P_4w_{21}^{+}-P_2w_{21}^{-}) \ and \
		J_{f}=2(P_4w_{2}^{+}-P_3w_{2}^{-})$. 
		%\begin{align}
		%J_{a}&=2(P_1k_{11}^{+}-P_2k_{11}^{-} )\label{eq:ja}\\
		%J_{b}&=2(P_2w_{1}^{+}-P_1w_{1}^{-})\label{eq:jb}\\
		%J_{c}&=2(P_2k_{12}^{+}-P_3k_{12}^{-}) \label{eq:jc}\\
		%J_{d}&=2(P_3k_{22}^{+}-P_4k_{22}^{-}) \label{eq:jd}\\
		%J_{e}&=2(P_4w_{21}^{+}-P_2w_{21}^{-}) \label{eq:je}\\
		%J_{f}&=2(P_4w_{2}^{+}-P_3w_{2}^{-}) \label{eq:jf}
		%\end{align}
		From equation \eqref{eq:j}, $J_a=\frac{(w_{1}^{+}k_{11}^{+}-w_{1}^{-}k_{11}^{-})K}{C_1}.$ Comparing, we get 
		$P_1=\frac{w_{1}^{+} K}{2C_1}$ and $P_2=\frac{w_{1}^{-} K}{2C_1}$. If $w_{1}^{+}=w_{1}^{-}=1$,  then $P_1=P_2=\frac{K}{2C_1}$. Hence, $J_b=\frac{(w_{1}^{-}w_{1}^{+}-w_{1}^{+}w_{1}^{-})K}{2C_1} =0$ and $J_I=(J_a+J_b)=J_a=\frac{(w_{1}^{+}k_{11}^{+}-w_{1}^{-}k_{11}^{-})K}{C_1}.$\\ 
		
		\noindent To calculate the other stationary currents $J_c, J_d, J_e$ and $J_f$, we use the Master equation of \eqref{eq:me} as,
		\begin{equation}
			\frac{d}{dt}\textbf{P}=\textbf{L}\textbf{P},
		\end{equation}
		where $\textbf{P}=\textbf{P}(t)=[P_1(t),P_2(t),P_3(t),P_4(t)]^{T}$ and $\textbf{L}$ is the stochastic transition matrix. \\
		
\begin{figure}[H]
	\centering
	\includegraphics[height=5cm,width=8cm,angle=0]{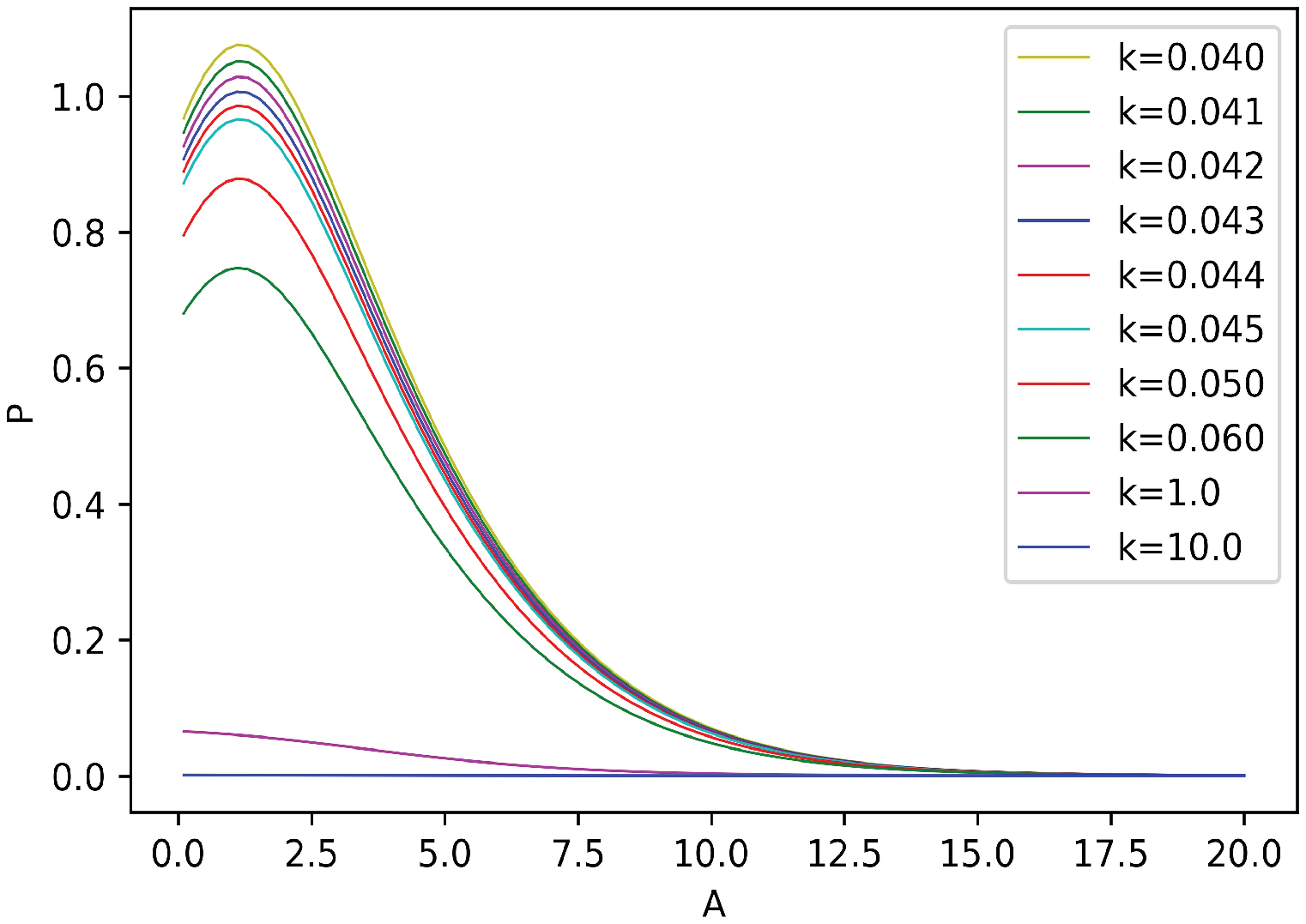}
	\caption{\small \textbf{Variation of the stationary probability distribution $P$ with affinity $\mathcal{A}$ at different values of the parameter $k$ for the interlinked two Rab species model system:} 
		We observe that $0\leq P\leq 1$ for $k\geq 0.044$. The rate constants at the interlinks $c$ and $e$ in Figure \ref{fig:model} are functions of the constant parameter $k$, viz. $k_{12}^{\pm}=w_{21}^{\pm}=f(k)$. This shows that the system can choose a range of constant $k$ values and hence a range of rate constant values $k_{12}^{\pm}$ and $w_{21}^{\pm}$ at the interlinks of the cascade. This implies a tunable range of rate constants.}
	\label{fig:probdis}
\end{figure}

		\noindent The stationary probability vector $\textbf{P}^{st}=[P_1,P_2,P_3,P_4]^{T}$ is obtained from $\frac{d}{dt}\textbf{P}^{st}=\textbf{L}^a(z=0)\ \textbf{P}^{st}=0$ as,
		\begin{equation}
			\begin{bmatrix}
				$$-(k_{11}^{+}+w_{1}^{-})$$ & $$(k_{11}^{+}+w_{1}^{-})$$ & $0$ & $0$ \\
				$$(k_{11}^{-}+w_{1}^{+})$$ & $$-(k_{11}^{-}+w_{1}^{+}+k_{12}^{+}+w_{21}^{-})$$ & $$k_{12}^{+}$$ &$$w_{21}^{-}$$ \\
				$$0$$ & $$k_{12}^{-}$$ & $$-(k_{12}^{-}+w_{2}^{-}+k_{22}^{+})$$ &$$(k_{22}^{+}+w_{2}^{-})$$ \\
				$$0$$ & $$w_{21}^{+}$$ & $$(k_{22}^{-}+w_{2}^{+})$$ &$$-(k_{22}^{-}+w_{2}^{+}+w_{21}^{+})$$ 
			\end{bmatrix}
			\begin{bmatrix}
				$$P_1$$\\
				$$P_2$$ \\
				$$P_3$$ \\
				$$P_4$$ 
			\end{bmatrix}=0
		\end{equation}
		We get the following equations as,
		\begin{align}
			&P_1(k_{11}^{+}+w_{1}^{-})=P_2(k_{11}^{+}+w_{1}^{-}) \label{eq:pp1}\\
			&(k_{11}^{-}+w_{1}^{+})P_1-(k_{11}^{-}+w_{1}^{+}+k_{12}^{+}+w_{21}^{-})P_2+k_{12}^{+}P_3+w_{21}^{-}P_4=0  \label{eq:pp2}\\
			&k_{12}^{-}P_2-(k_{12}^{-}+w_{2}^{-}+k_{22}^{+})P_3+(k_{22}^{+}+w_{2}^{-})P_4=0  \label{eq:pp3}\\
			&w_{21}^{+}P_2+(k_{22}^{-}+w_{2}^{+})P_3-(k_{22}^{-}+w_{2}^{+}+w_{21}^{+})P_4=0 \label{eq:pp4} 
		\end{align}
		{\noindent}From equation $\eqref{eq:pp1}$, we get $P_1=P_2$. Substituting this in equation $\eqref{eq:pp2}$, we get $P_2=\frac{k_{12}^{+}P_3+w_{21}^{-}P_4}{k_{12}^{+}+w_{21}^{-}}$. Again substituting this $P_2$ expression in equation $\eqref{eq:pp3}$, we get,
		\begin{align}
			&\left(\frac{k_{12}^{-}k_{12}^{+}P_3+w_{21}^{-}k_{12}^{-}P_4}{k_{12}^{-}+w_{21}^{-}}\right)-(k_{12}^{-}+w_{2}^{-}+k_{22}^{+})P_3+(k_{22}^{+}+w_{2}^{-})P_4 =0  \nonumber\\
			&\Rightarrow P_3\left(\frac{k_{12}^{-}k_{12}^{+}}{k_{12}^{-}+w_{21}^{-}}-(k_{12}^{-}+w_{2}^{-}+k_{22}^{+})\right)+P_4\left(\frac{w_{21}^{-}k_{12}^{-}}{w_{21}^{-}+k_{12}^{+}}+(k_{22}^{+}+w_{2}^{-})\right) =0  \nonumber \\
			&\Rightarrow P_3\left[k_{12}^{-}k_{12}^{+}-(k_{12}^{-}+w_{2}^{-}+k_{22}^{+})(w_{21}^{-}+k_{12}^{+})\right]+P_4\left[w_{21}^{-}k_{12}^{-}+(k_{22}^{+}+w_{2}^{-})(w_{21}^{-}+k_{12}^{+})\right] =0 \nonumber  \\
			&\Rightarrow P_3=P_4. \nonumber 
		\end{align}
		Putting $P_3=P_4$ in equation \eqref{eq:pp4}, we get,
		\begin{align}
			& w_{21}^{+}P_2+(k_{22}^{-}+w_{2}^{+})P_4-(k_{22}^{-}+w_{2}^{+}+w_{21}^{+})P_4 = 0\nonumber\\
			&\Rightarrow P_2= P_4  \nonumber 
		\end{align} Let $ P_1=P_2=P_3=P_4=P$. Then,
		\begin{equation}
			P= \frac{K}{2C_1}=P(k,\mathcal{A})  \label{eq:ps}
		\end{equation}
		We numerically calculate the values of the stationary probability $P$ of equation \eqref{eq:ps} and plot the variation of $P$ with the affinity $\mathcal{A}$ for different values of $k$, as shown in Figure \ref{fig:probdis}. We see that there is a range of the constant parameter $k$ allowed for our double Rab species interlinked model. We find that $0<P\leq 1$ for $k \geq 0.044$.\\
		
		\noindent Substituting the expressions of the parameters, we calculate  $J_{c}=\left(ke^{3\mathcal{A}/4}-\frac{1}{k}\right)\frac{K}{C_1}$; $J_{d}=\left(e^{\mathcal{A}/4}-1\right)\frac{K}{C_1};\\
		J_{e}=\left(\frac{e^{3\mathcal{A}/4}}{k}-k\right)\frac{K}{C_1} \ and \ J_{f}=0$.
		%\begin{align}
		%J_{c}&=\left(ke^{3\mathcal{A}/4}-\frac{1}{k}\right)\frac{K}{C_1} \\
		%J_{d}&=\left(e^{\mathcal{A}/4}-1\right)\frac{K}{C_1} \\
		%J_{e}&=\left(\frac{e^{3\mathcal{A}/4}}{k}-k\right)\frac{K}{C_1} \\
		%J_{f}&=0
		%\end{align}
		Hence, we get,
		\begin{align}
			J_{II}&=J_c+J_d+J_e=\frac{K}{C_1}\left(ke^{3\mathcal{A}/4}-k+e^{\mathcal{A}/4}-\frac{1}{k}+\frac{e^{3\mathcal{A}/4}}{k}-1\right)\\
			J_{III}&=J_d+J_f=\left(e^{\mathcal{A}/4}-1\right)\frac{K}{C_1}
		\end{align}
		The thermodynamic cost of equation \eqref{eq:finale} becomes,
		\begin{align}
			Q_a &= A_{I}\bigg[\coth{(\mathcal{A}/8)}-\left(\frac{LC_1-KC_2}{C_{1}^{2}}\right) 2(e^{\mathcal{A}/4}-1) \bigg ] \nonumber \\
			& + \frac{2D_a }{J_{a}^2} \Bigg[\frac{K}{C_1}\left(ke^{3\mathcal{A}/4}-k+e^{\mathcal{A}/4}-\frac{1}{k}+\frac{e^{3\mathcal{A}/4}}{k}-1\right)A_{II}+\left(e^{\mathcal{A}/4}-1\right)\frac{K}{C_1}A_{III}\Bigg].  \label{eq:qfinal}
		\end{align}
		Using equations \eqref{eq:j1} and \eqref{eq:diff}, we calculate the term $\frac{2D_a}{J_{a}^{2}}$ as,
		\begin{align}
			\frac{2D_a}{J_{a}^{2}}
			%\nonumber\\
			%&=2\frac{-(C_{0}^{''}-2C_{1}^{'}J_a-2C_2{J_{a}^{2}})}{2C_1\frac{(-C_0')^2}{C_1^2}} \nonumber \\
			&=(-C_{0}^{''}+2C_{1}^{'}J_a+2J_{a}^{2}C_2)\frac{C_1}{(C_0')^2} \nonumber \\
			%&=\bigg [(k_{11}^{+}w_1^{+}+k_{11}^{-}w_1^{-})K+2(-k_{11}^{+}w_1^{+}+k_{11}^{-}w_1^{-})L(k_{11}^{+}w_1^{+}-k_{11}^{-}w_1^{-})\frac{K}{C_1}+2(k_{11}^{+}w_1^{+}-k_{11}^{-}w_1^{-})^2K^2\frac{C_2}{C_{1}^{2}}\bigg ]\frac{C_1}{C_{0}^{'2}}\nonumber \\
			&=\bigg [(k_{11}^{+}w_1^{+}+k_{11}^{-}w_1^{-})K-2(k_{11}^{+}w_1^{+}-k_{11}^{-}w_1^{-})^2\frac{LK}{C_1}+2(k_{11}^{+}w_1^{+}-k_{11}^{-}w_1^{-})^2K^2\frac{C_2}{C_{1}^{2}}\bigg ]\frac{C_1}{C_{0}^{'2}}\nonumber \\
			&=\bigg [(k_{11}^{+}w_1^{+}+k_{11}^{-}w_1^{-})K-\left(\frac{L}{C_1}-\frac{C_2K}{C_1^{2}}\right)2(k_{11}^{+}w_1^{+}-k_{11}^{-}w_1^{-})^2K \bigg ]\frac{C_1}{C_{0}^{'2}}\nonumber \\
			&= \bigg [(e^{\mathcal{A}/4}+1)K-\left(\frac{LC_1-C_2K}{C_1^{2}}\right)2(e^{\mathcal{A}/4}-1)^2K \bigg ]\frac{C_1}{C_{0}^{'2}}.\label{eq:term}
		\end{align}
		%\begin{eqnarray}
		%Q_a&=& A_{I}\bigg[\coth{(\mathcal{A}/8)} +\left(\frac{LC_1-KC_2}{C_{1}^{2}}\right) 2(e^{(\mathcal{A}/4)}-1) \bigg ] \nonumber \\
		%&&+ \bigg [(e^{\mathcal{A}/4}+1)K+\left(\frac{LC_1-C_2K}{C_1^{2}}\right)2(e^{\mathcal{A}/4}-1)^2K \bigg ]\times \frac{C_1}{C_{0}^{'2}}\nonumber \\
		%&& \times \Bigg[\frac{K}{C_1}\left(ke^{3\mathcal{A}/4}-k+e^{\mathcal{A}/4}-\frac{1}{k}+\frac{e^{3\mathcal{A}/4}}{k}-1\right)A_{II}\nonumber \\
		%&&+\left(e^{\mathcal{A}/4}-1\right)\frac{K}{C_1}A_{III}\Bigg] \nonumber
		%&=& A_{I}\bigg[\coth{(\mathcal{A}/8)} +\left(\frac{LC_1-KC_2}{C_{1}^{2}}\right) 2(e^{(\mathcal{A}/4)}-1) \bigg ] \nonumber \\
		%&&+ \bigg [(e^{\mathcal{A}/4}+1)K+\left(\frac{LC_1-C_2K}{C_1^{2}}\right)2(e^{\mathcal{A}/4}-1)^2K \bigg ]\times \frac{1}{(-e^{\mathcal{A}/4}+1)^2K}\nonumber\\
		%&&\times \Bigg[\left(ke^{3\mathcal{A}/4}-k+e^{\mathcal{A}/4}-\frac{1}{k}+\frac{e^{3\mathcal{A}/4}}{k}-1\right)A_{II}+\left(e^{\mathcal{A}/4}-1\right)\mathcal A_{III}\Bigg] \nonumber 
		%\end{eqnarray}
		Putting equation \eqref{eq:term} in \eqref{eq:qfinal}, we get,
		\begin{align}
			Q_{a}&= A_{I}\bigg[\coth{(\mathcal{A}/8)}-\left(\frac{LC_1-KC_2}{C_{1}^{2}}\right) 2(e^{(\mathcal{A}/4)}-1) \bigg ] \nonumber \\
			&+ \bigg [(e^{\mathcal{A}/4}+1)K-\left(\frac{LC_1-C_2K}{C_1^{2}}\right)2(e^{\mathcal{A}/4}-1)^2K \bigg ] \frac{C_1}{C_{0}^{'2}}\nonumber \\
			&\times \Bigg[\frac{K}{C_1}\left(ke^{3\mathcal{A}/4}-k+e^{\mathcal{A}/4}-\frac{1}{k}+\frac{e^{3\mathcal{A}/4}}{k}-1\right)A_{II}+\left(e^{\mathcal{A}/4}-1\right)\frac{K}{C_1}A_{III}\Bigg] \label{eq:qfull}\\
			&=Q_{a}^{ full} \nonumber 
		\end{align}
		We use the notation $Q_a^{full}$ to denote that all the affinities $A_I, A_{II}$ and $A_{III}$ of the three cycles are present in the equation. Now we calculate the constants $L, K, C_1, C_2$ present in the above expression of $Q_a^{full}$ as,
		\begin{align}
			L&= 3+\frac{e^{3\mathcal{A}/4}}{k}+\frac{1}{k} +e^{\mathcal{A}/4} \label{eq:ll}\\
			K&=\frac{2}{k}+\frac{e^{\mathcal{A}/4}}{k}+\frac{e^{3\mathcal{A}/4}}{k^2} +\frac{e^{\mathcal{A}}}{k} \label{eq:kk}\\
			C_1&=\frac{6}{k}+\frac{2e^{3\mathcal{A}/4}}{k}+\frac{3e^{3\mathcal{A}/4}}{k^2} +\frac{e^{\mathcal{A}}}{k^2}+\frac{e^{5\mathcal{A}/4}}{k}+\frac{4e^{\mathcal{A}}}{k}\frac{2e^{\mathcal{A}/4}}{k} + 4ke^{\mathcal{A}}\nonumber \\
			& + ke^{5\mathcal{A}/4}+4ke^{\mathcal{A}/4} + e^{\mathcal{A}/4}+ke^{\mathcal{A}/2} + 3ke^{3\mathcal{A}/4}+ e^{3\mathcal{A}/2}+3k +1    \label{eq:cone}\\
			C_2&=\frac{6}{k}+\frac{e^{3\mathcal{A}/4}}{k^2}+\frac{2e^{\mathcal{A}}}{k} +\frac{4e^{3\mathcal{A}/4}}{k}+1+2ke^{\mathcal{A}} + 4ke^{3\mathcal{A}/4}+e^{3\mathcal{A}/2} \nonumber \\
			&+2k e^{\mathcal{A}/4}+4k+\frac{e^{\mathcal{A}/4}}{k} +e^{\mathcal{A}/2}+ 6e^{\mathcal{A}/4}\label{eq:ctwo}
		\end{align}
		We now analyse the equation \eqref{eq:qfull} in two different cases. First, we consider that only the affinity in the first cycle is non-zero, while that in the remaining two cycles are zeroes. Second, we consider that all the affinities in the three cycles are non-zeroes. The reason for studying these two cases is to investigate the thermodynamic costs for different situations of the interlinking mechanism.\\
		
		\noindent \textbf{Analytical results of Case-I:} In equation \eqref{eq:qfull}, we take the affinities in the three cycles as $A_I=\mathcal{A}\neq 0$ and $A_{II}=A_{III}=0$. This means that the reactions in the second and third cycles are in equilibrium. Dividing by the number of states $(n=4)$, we get the thermodynamic cost of equation \eqref{eq:qfull} as, 
		\begin{align}
			\frac{Q_a}{4}&=\frac{\mathcal{A}}{4}\bigg[\coth{(\mathcal{A}/8)}-\left(\frac{LC_1-KC_2}{C_{1}^{2}}\right) 2(e^{\mathcal{A}/4}-1) \bigg ]=\frac{Q_a^{half}}{4}. \label{eq:qhalf}
		\end{align} 
		The notation $Q_a^{half}$ implies that only the first term corresponding to $A_I=\mathcal{A}\neq 0$ is present in the equation \eqref{eq:qfull}. As the value of $k \rightarrow large$, then $L \sim constant$, $K \sim very \ small$, $C_1\sim large $ and $C_2\sim large $ (from equations \eqref{eq:ll}, \eqref{eq:kk},\eqref{eq:cone}, \eqref{eq:ctwo}). Hence, $\frac{Q_a^{half}}{4} \rightarrow \frac{\mathcal{A}}{4} \coth{(\mathcal{A}/8)}$ as $k \rightarrow large$. \\ 
		
\begin{figure}[H]
	\centering
	\includegraphics[height=18cm,width=14cm,angle=0]{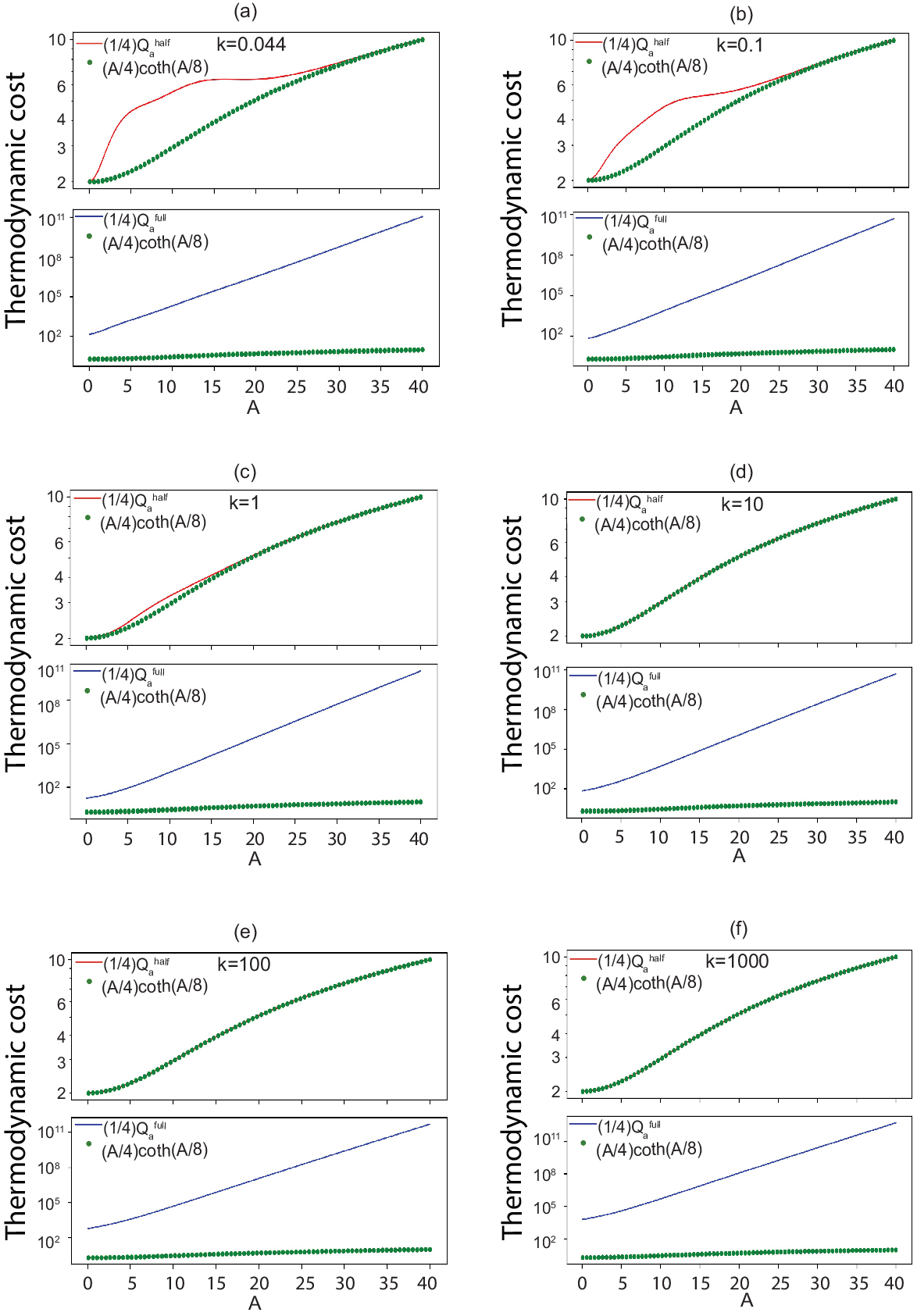}
	\caption{\textbf{Numerical results of the interlinked two Rab species model}: In each sub-plot, the upper panels are the numerical results corresponding to the analytical calculations of \textbf{Case-I} and the lower panels are those of \textbf{Case-II}. We see that for the interlinked system with an affinity present in the first cycle and all the other affinities in the remaining cycles are zeroes, $(Q_a^{half}/4) \rightarrow 2$ as the affinity $\mathcal{A}\rightarrow 0$ for all values of $k$.  Again, when all the cycles are in non-equilibrium, the $(Q_a^{full}/4)$ is large for all values of $k$. Hence, the Thermodynamic Uncertainty Relation for the interlinked two Rab species model is proved as $(Q_a^{half}/4)\geq 2$ and $(Q_a^{full}/4)\geq 2$ for all values of $\mathcal{A}$. For a given $k$ value, $(Q_a^{full}/4)\gg (Q_a^{half}/4)$ at all $\mathcal{A}$ values. These results show that the mechanism of interlinking at non-equilibrium enhances the thermodynamic costs.}
	\label{fig:qanalysis}
\end{figure}

		\noindent \textbf{Numerical results of Case-I:} We now numerically calculate the term $\frac{Q_a^{half}}{4}$ of equation \eqref{eq:qhalf}. The upper panels of each subplot of Figure \ref{fig:qanalysis} represent the numerical calculations of the thermodynamic cost of equation \eqref{eq:qhalf} along with the curve of $\frac{\mathcal{A}}{4} \coth{(\mathcal{A}/8)}$ for different values of $k$. From the upper panels, we see that the term $\frac{Q_a^{half}}{4}\rightarrow 2$ as $\mathcal{A}\rightarrow 0$ and $\frac{Q_a^{half}}{4}>2$ as $\mathcal{A}$ increases, for all values of $k$.  Hence, we prove the Thermodynamic Uncertainty Relation that the thermodynamic cost $\frac{Q_a^{half}}{4}\geq 2$ for the interlinking condition where only affinity $A_I=\mathcal{A}$ is present and all the other affinities are zero (equilibrium). From the results, we observe that the curve of $\frac{Q_a^{half}}{4}$ coincides with the curve of $\frac{\mathcal{A}}{4} \coth(\mathcal{A}/8)$ as $k$ increases.  This observation is explained by the above analytical results of Case-I. We also observe that the thermodynamic cost is less at a given affinity $(\sim 10$ at $\mathcal{A}=40)$ for all values of $k$ which implies less precision. \\ 
		
		{\noindent}\textbf{Analytical results of Case-II:} We know investigate the thermodynamic cost when affinities in all the three cycles are non-zero (far from equilibrium). In equation \eqref{eq:qfull}, we take the same affinity in the first, second and third cycles as $A_{I}=A_{II}=A_{III}=\mathcal{A}\neq 0.$\\
		Equation \eqref{eq:qfull} becomes,
		\begin{align}
			\frac{Q_a^{full}}{4}&= \frac{\mathcal{A}}{4}\bigg[\coth{(\mathcal{A}/8)}-\left(\frac{LC_1-KC_2}{C_{1}^{2}}\right) 2(e^{\mathcal{A}/4}-1) \bigg ] \nonumber \\
			&+ \frac{1}{4}\times \bigg [(e^{\mathcal{A}/4}+1)K-\left(\frac{LC_1-C_2K}{C_1^{2}}\right)2(e^{\mathcal{A}/4}-1)^2K \bigg ]\times \frac{1}{(-e^{\mathcal{A}/4}+1)^2K}\nonumber\\
			&\times \Bigg[\left(ke^{3\mathcal{A}/4}-k+e^{\mathcal{A}/4}-\frac{1}{k}+\frac{e^{3\mathcal{A}/4}}{k}-1\right)\mathcal{A}+\left(e^{\mathcal{A}/4}-1\right)\mathcal{A}\Bigg] \nonumber \\
			&= \frac{\mathcal{A}}{4}\bigg[\coth{(\mathcal{A}/8)}-\left(\frac{LC_1-KC_2}{C_{1}^{2}}\right) 2(e^{\mathcal{A}/4}-1) \bigg ] \nonumber \\
			&+\frac{\mathcal{A}}{4} \times \bigg [\frac{(e^{\mathcal{A}/4}+1)}{(-e^{\mathcal{A}/4}+1)^2}-\left(\frac{LC_1-C_2K}{C_1^{2}}\right)2\bigg ] \times \left(ke^{3\mathcal{A}/4}-k+2e^{\mathcal{A}/4}-\frac{1}{k}+\frac{e^{3\mathcal{A}/4}}{k}-2\right)  \nonumber \\
			&=\left(\frac{Q_a^{half}}{4}\right)+ \left(\frac{Q_a^{extra}}{4}\right). \label{eq:final}
		\end{align}
		We have
		\begin{equation}
			\left(\frac{Q_a^{half}}{4}\right)=\frac{\mathcal{A}}{4}\coth{(\mathcal{A}/8)}-\frac{\mathcal{A}}{4}\bigg[\left(\frac{LC_1-KC_2}{C_{1}^{2}}\right) 2(e^{(\mathcal{A}/4)}-1) \bigg ]. \label{eq:qhalf1}
		\end{equation}
		\noindent Equation \eqref{eq:qhalf1} is the same as that of equation \eqref{eq:qhalf}. From the analytical and numerical results of \textbf{Case-I}, we have seen that $\frac{Q_a^{half}}{4} \rightarrow 2$ as $ \mathcal{A} \rightarrow 0$ and $\frac{Q_a^{half}}{4} \rightarrow \frac{\mathcal{A}}{4} \coth{(\mathcal{A}/8)}$ as $k \rightarrow large$.\\
		
		\noindent We have 
		\begin{align}
			&\left(\frac{Q_a^{extra}}{4}\right)=\frac{\mathcal{A}}{4} \bigg [\frac{(e^{\mathcal{A}/4}+1)}{(-e^{\mathcal{A}/4}+1)^2}-\left(\frac{LC_1-C_2K}{C_1^{2}}\right)2\bigg ] \times \left(ke^{3\mathcal{A}/4}-k+2e^{\mathcal{A}/4}-\frac{1}{k}+\frac{e^{3\mathcal{A}/4}}{k}-2\right). \label{eq:extra}
		\end{align}
		\noindent As $k \rightarrow large$, the $L \sim constant$, $K \sim very \ small$, $C_1\sim large $ and $C_2\sim large $ (from equations \eqref{eq:ll}, \eqref{eq:kk},\eqref{eq:cone}, \eqref{eq:ctwo}). Hence, $\frac{Q_a^{extra}}{4}\rightarrow large$ as $k \rightarrow large$.\\
		
		\noindent From equation \eqref{eq:final}, $\frac{Q_a^{full}}{4}\rightarrow large $ as $k \rightarrow large$ for all values of $\mathcal{A}$.  \\ 
		
		\noindent Since equation \eqref{eq:qhalf1} is the same as that of equation \eqref{eq:qhalf}, we can now investigate the effect on the thermodynamic cost when we add the extra term of \eqref{eq:extra} to equation \eqref{eq:qhalf} of the previous analysis of Case-I. That is, we analyse the thermodynamic cost when the interlinking mechanism has non-equilibrium processes.\\
		
		\noindent \textbf{Numerical results of Case-II:} We calculate the thermodynamic cost of equation \eqref{eq:final}. We plot the variation of $\frac{Q_a^{full}}{4}$ with affinity $\mathcal{A}$ for different values of $k$ in the lower panels of each subplots of Figure \ref{fig:qanalysis}. We observe that $\frac{Q_a^{full}}{4} >2$ and $\frac{Q_a^{full}}{4} \rightarrow large$ for all values of $\mathcal{A}$ and $k$. Hence, we prove the Thermodynamic Uncertainty Relation that the thermodynamic cost $\frac{Q_a^{full}}{4}\geq 2$ for the interlinking condition where all affinities are non-zero. We see that $\frac{Q_a^{full}}{4}\sim 10^{11}$ at $\mathcal{A}\sim 40$ for all values of $k$. \\
		
\begin{figure}[H]
	\centering
	\includegraphics[height=16cm,width=14cm,angle=0]{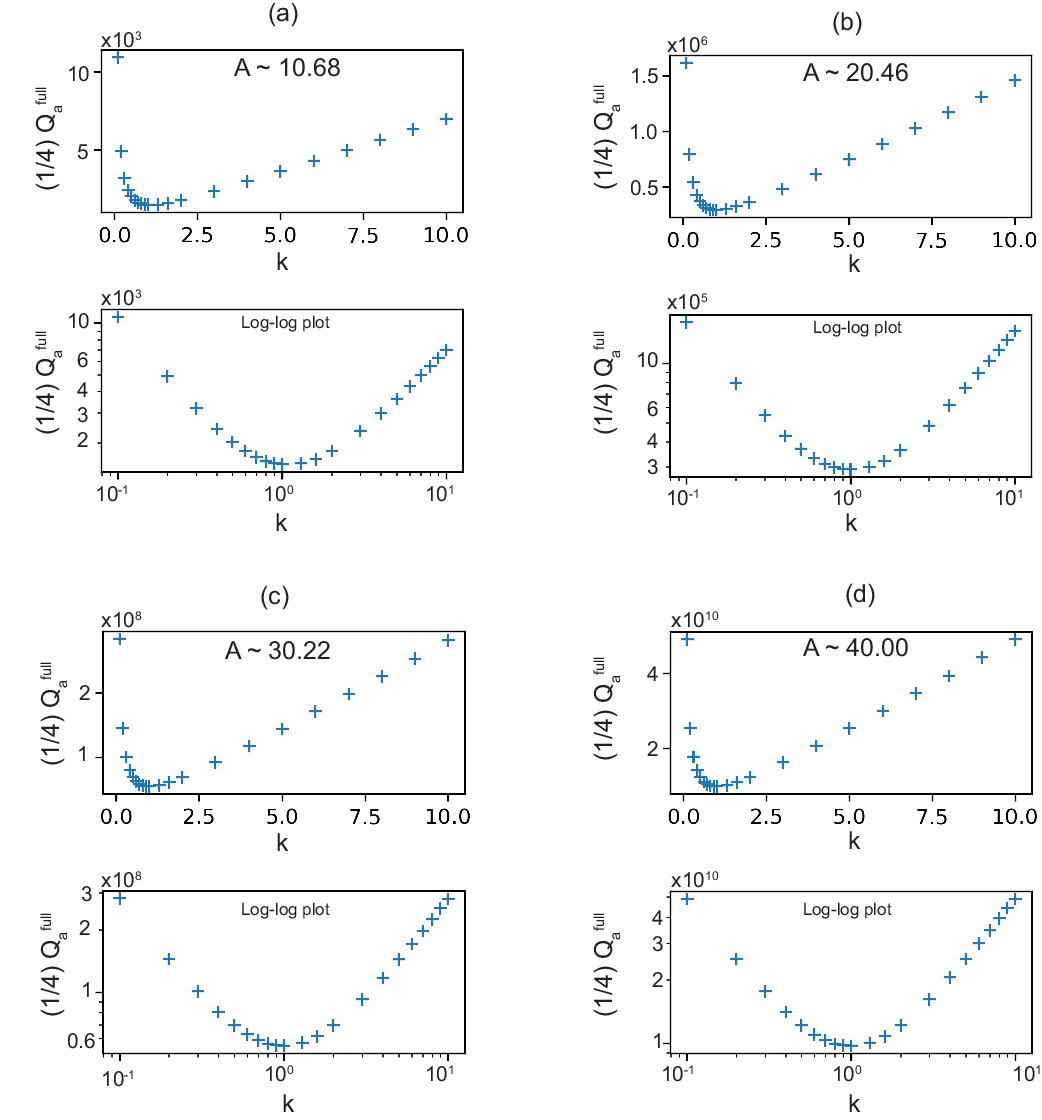}
	\caption{\textbf{Variation of $(Q_a^{full}/4)$ with respect to the parameter $k$ at different values of affinity $\mathcal{A}$:} In each sub-plot,  the upper panels are in linear scale, whereas the lower panels are in log-log plots.  In each subfigure, we see that $(Q_a^{full}/4)$ decreases sharply, reaches a minima and then increases gradually with an increase in the value of the parameter $k$. This shows that at small and large values of $k$, the thermodynamic cost is large. This implies that the interlinked system chooses a range of $k$ values to optimise the thermodynamic cost and hence precision. }
	\label{fig:range}	
\end{figure}

		\noindent From Figure \ref{fig:range}, we observe that, for a given affinity $\mathcal{A}$, $\frac{Q_a^{full}}{4}$ first decreases sharply and then increases gradually as $k$ increases. This shows a minima of $\frac{Q_a^{full}}{4}$. This can be explained with equation \eqref{eq:final}. The term $\frac{Q_a^{half}}{4}$ decreases as $k$ increases and goes to a constant minimum value for all values of $k$. On the otherhand, the term $\frac{Q_a^{extra}}{4}$ keeps increasing as $k$ increases. Hence, the total thermodynamic cost $\frac{Q_a^{full}}{4}$ grows large for further increase in $k$. \\ 
		
		\noindent We now summarise the results from \textbf{Case-I} and \textbf{Case-II}. First, we prove the TUR in both the models of single Rab species switching and interlinked cascade of two Rab species. We observe that when the two Rab species are interlinked with a non-zero affinity in the upstream cycle and all the affinities in the downstream cycles being zero, the thermodynamic cost is only $\sim 10$ at a given affinity $\mathcal{A}=40$. The TUR shows that this less thermodynamic cost implies less precision of biomolecular processes. However, when the two species are interlinked with non-zero affinities in all the cycles, the thermodynamic cost is greatly enhanced to $\sim 10^{11}$ at the given $\mathcal{A}=40.$ Our results thus show that when the two Rab species are interlinked and are at far from equilibrium, the thermodynamic cost is significantly increased, and hence this leads to a higher precision of the performance of biomolecular processes of Rab proteins.\\ 
		
		\noindent We now give a short analysis of the diffusion coefficient formula. When we use the formula \eqref{eq:bs} given by Barato and Seifert, we find that, for our network model, the term $\frac{Q_{a}^{half}}{4} \geq 2$ for $k\geq 1$. For lesser values of $k$, $\frac{Q_{a}^{half}}{4} < 2$. Again, when we use the formula \eqref{eq:koza} given by Koza, we find similar behaviour of the graphs as in Figure \ref{fig:qanalysis} but in the IV quadrant. This means that the thermodynamic costs have the same magnitudes but negative signs. We thus take the negative of the original formula \eqref{eq:koza} given by Koza and use the equation \eqref{eq:diff} in all our analyses. 
		\section{Conclusion}
		\label{sec:5.5}
		\noindent To conclude, we investigate the Thermodynamic Uncertainty Relation (TUR) in a non-equilibrium biological model of interlinked cascades of RabGTPase proteins. For a single Rab species switching model, we prove the TUR that the thermodynamic cost $Q\rightarrow 2$ (minimum) as the affinity $\mathcal{A} \rightarrow 0$. We also find that $Q\rightarrow 2$ as $\mathcal{A} \rightarrow large$. This implies that the thermodynamic cost is minimised (or precision is lessened) at far from equilibrium ($\mathcal{A} \rightarrow large$), which is not desirable. We now interlink the Rab species with another Rab species and investigate the thermodynamic cost. For the interlinked two Rab species model, we again prove the TUR. When all the affinities in the cycles are zeroes except for the upstream cycle, we find that the thermodynamic cost is less, which is again not desirable due to less precision. However, the thermodynamic cost is greatly enhanced when the two Rab species are interlinked with all non-zero affinities. This shows that the Rab proteins try to optimise the precision of their biomolecular processes by forming interlinks at far from equilibrium. Our results show the significance of interlinked cascades in the biomolecular processes of RabGTPase proteins. We also find that except at some particular value of the constant parameter $k$ where the thermodynamic cost $Q$ is minimal, the $Q$ is large at small and large values of $k$. This shows a range of tunable rate constants involved in the reaction channels of the models. Our results show that the energetics and cost of such a non-equilibrium reactions system are significantly controlled by the reaction rates involved in the underlying reactions network, as also pointed out in \cite{copy}. \\

		{\noindent}Rab GTPase proteins of our interlinked model system are involved in many important biochemical pathways. They are the major regulators in cellular membrane trafficking, such as vesicle formation, transport, tethering and fusion in eukaryotic cells \cite{chia}. With nearly seventy members of the Rab family in human beings, Rab GTPases regulate many functions such as cell proliferation, cell migration, and cell metabolism. Hence, the impairment of pathways involved with Rab GTPases is connected with many diseases \cite{maria,li, muller}. In humans, impairments related to Rab GTPases and their associated regulatory proteins cause malignancies such as Griscelli syndrome, Charcot–Marie–Tooth disease, kidney disease, vascular disease, thyroid disease and choroideremia \cite{stein,guadagno}. Overexpression of several members of the Rab family is found in various cancer tissues, including breast, liver, prostate, lung and oral \cite{subramani,chia,stein,chen,romano,yang,tzeng,oral}. Rab25 is related to tumour cell migration and invasion of epithelial cancers \cite{chia}. Rab13 is a potential driver of cancer progression \cite{maria}. There have been recent target studies on Rabs to understand how the dysregulation of their associated functions leads to disorders including cancer and to find potential therapeutic strategies \cite{guadagno, xin}. Our results from the perspectives of the Thermodynamic Uncertainty Relation enhance our understanding that Rab GTPases optimise the thermodynamic cost and precision of their biomolecular processes by forming interlinks at  far from equilibrium. The interplay between the cost and precision by manipulating a tunable range of rate constants can regulate the biomolecular processes to prevent dysfunctioning leading to disorders.\\ 
		
\begin{figure}[H]
	\centering
	%\vspace{-2cm}
	\includegraphics[height=7cm,width=11cm]{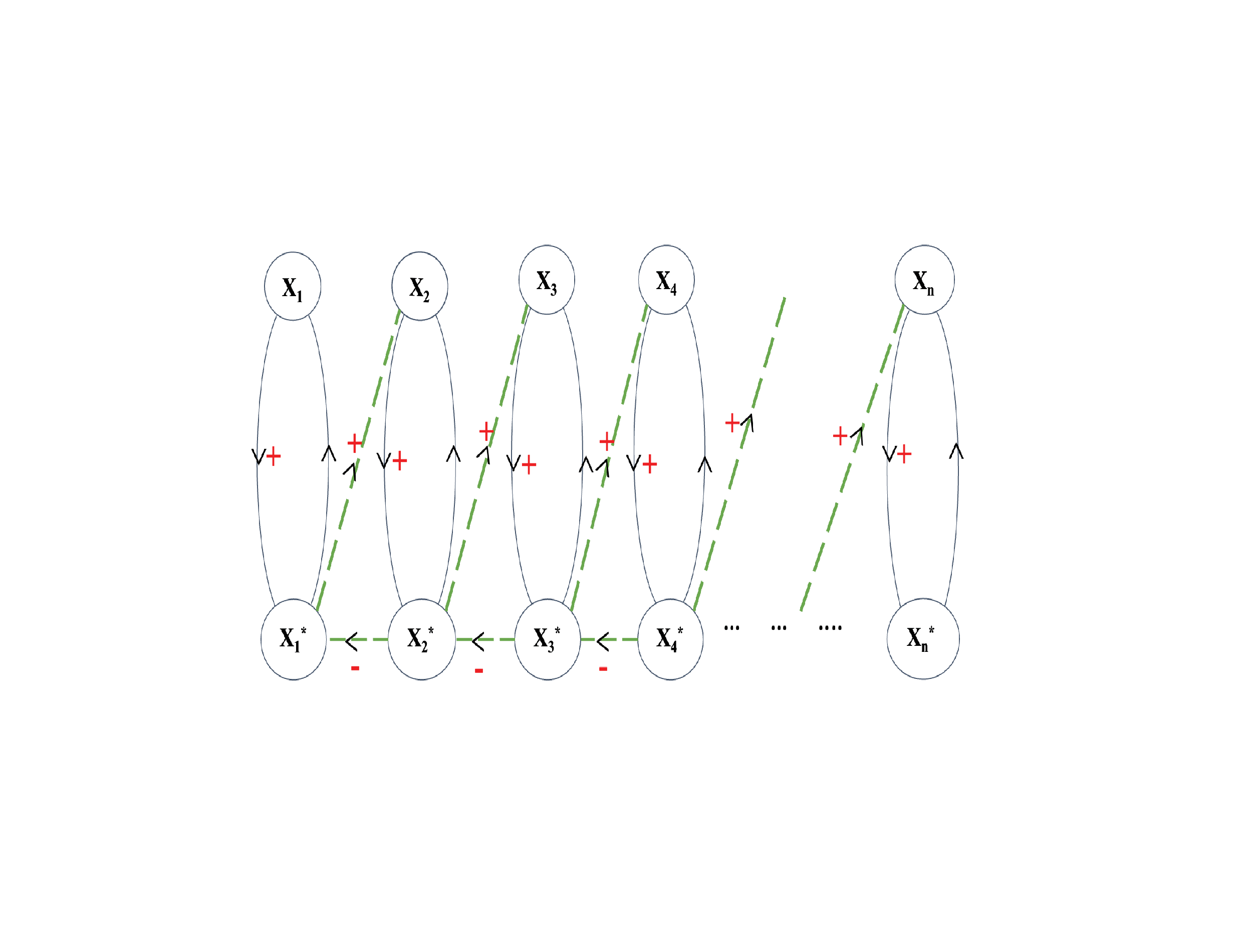}
	\vspace{-1.5cm}
	\caption{ \small \textbf{Modelling of the Rab GEF and GAP cascades using a general network of Markov states: } $X_1, X_2,\dots, X_n$ denote the inactive-GDP bound states of the RabGTPases, whereas $X_1,^{*} X_2^{*},\dots, X_n^{*}$ denote the active GTP-bound states of the RabGTPases. The green dashed lines indicate the interlinking between any two different RabGTPase species.  The $+$ signs in red colour indicate an activation or positive feedback and the $-$ signs in red colour indicate a deactivation or negative feedback in the direction of the arrows. We model the interlinked Rab cascades with an arrangement of triangular network motifs with Positive-Positive-Negative feedbacks.}
	\label{fig:representation}
\end{figure}

			{\noindent}In cancer, biological cellular networks such as metabolic networks or signalling networks are disregulated. Due to the huge network size and highly non-linear nature, it is difficult to study such complex biological networks with mathematical models and experiments. However, all such complex networks can be reduced in their basic building blocks called network motifs. By studying the motifs in terms of dynamics and functionality, we can enhance our understanding of cancer biology. To understand the underlying connectivity of small GTPases signalling,  a small GTPases protein interaction network is constructed using a systems-level approach based on experimentally validated interactions \cite{delprato}. Our network structure of the interlinked Rab GTPases cascade (Figure \ref{fig:representation}) comprises triangular motifs. Several Rab proteins are known to be involved in various cancer. Our analysis shows that triangular motifs are correlated with cellular networks of cancer. Indeed, the network biology approach finds triangular network motifs in many cancer studies \cite{ali,rocky,malik, andre,  cloutier, jeon,schramm}. Our study thus highlights a close relationship between the thermodynamic cost-precision, triangular motifs and cancer.  \\ 
		
		{\noindent}The interlinked Rab GTPases cascade (Figure \ref{fig:mechanism}) is a complex feedback control system, where not only the Rab proteins switch between their inactive GDP-bound state and their active GTP-bound state, but also the active upstream Rab protein promotes downstream signalling and the active downstream Rab protein, in turn, deactivates the active upstream Rab protein through the regulation of GEF and GAP \cite{krishnan}. In this way, Rab proteins control the intracellular transport in both spatial and timed manners \cite{jordens}. When two or more two Rab proteins are involved in the cascade, they function as oscillators \cite{ehrmann}. Our network structure (Figure \ref{fig:representation}) consists of triangular network motifs of a long negative (Positive-Positive-Negative) feedback. Coupled Positive-Negative feedback loops are important signal transduction motifs that allow cellular circuits to give proper rapid responses to external fluctuations and are robust to such fluctuations \cite{qian,kim1,anan,klinke,kim}. Such loops can achieve a wide range of tunable frequencies \cite{ferrell,tian1,tsai}. We expect the network motifs in Figure \ref{fig:representation} to be robust to fluctuations since our results show a range of tunable rate constants. Hence, our results show that the interlinked cascade or oscillator can achieve a range of tunable frequencies to maintain its robustness and optimise the precision of its performance. Not only the coherence of oscillations is improved by increasing the energy consumption, but also the coherence and oscillation period become robust to fluctuations in rates from the noisy environment of a cell \cite{junco}.   \\  
		
		\noindent Spontaneous pattern formation has been an interesting research area in the field of non-equilibrium processes since the groundbreaking theoretical works of Alan Turing on reaction-diffusion systems \cite{turing}. The dynamics of cell signaling happens in both spatial and temporal dimensions \cite{kholodenko}. The signaling network structure of Rho GTPases, another sub-family of Ras superfamily, enables spontaneous, self-limiting patterns of sub-cellular contractility by generating pulses and propagating waves of cell contractions in space and time \cite{graessl}. In biochemical oscillators and microtubule-kinesin active flow systems, a self-similarity in the underlying non-equilibrium reaction networks is required \cite{yu}. The relation between the thermodynamic cost and the precision of a spontaneous pattern is studied in the reaction-diffusion model of the Brusselator in 1-D space \cite{rana}. 
		Again, using mathematical modelling and in vitro reconstitution, the robustness of protein self-organization (patterning) in \textit{Escherichia coli} Min system is studied where it is found that interlinked functional switching of both Min CDE proteins, rather than one, imparts robustness in biological pattern-forming systems \cite{denk}. It will be interesting to observe in real experimental set ups the dynamic activity patterns of Rab GTPases. In this connection, our analysis implies that a self-organisation in the biochemical oscillator of Rab GTPases will emerge from the well-organised arrangement of self-similar triangular network motifs (Figure \ref{fig:representation}). It will be again interesting to study the cost-precision trade-off of the complex self-organised patterns arising at far from equilibrium for the interlinked Rab cascade. In this regards, our results show that interlinking of two Rab proteins at far from equilibrium maintains robustness and increases precision. 
		\section{Authors' contribution}
		{\noindent}The conceptualisation of the present work is done by ALC and RKBS. ALC carried out the analytical calculations, numerical analysis, and the preparation of associated figures. Both authors analysed the results, wrote, discussed and approved the final manuscript.
			\section{Competing financial interests}
				{\noindent}The authors declare no competing financial interests.	
			\section{Acknowledgements}
		\noindent ALC is an INSPIRE Fellow (DST/INSPIRE/03/2017/002925 with INSPIRE Code IF180043) and acknowledges the Department of Science and Technology (DST), Government of India for providing financial support (Order No: DST/INSPIRE Fellowship/2018/IF180043) under the INSPIRE program.  RKBS acknowledges DBT-COE, India, for providing financial support.

\vspace{-1cm}

%\begin{figure}[h]
%\includegraphics[height=7cm,width=10cm,angle=0]{Pall.png}
%\caption{\large \textit{Variation of Stationary Probability against Affinity }}
%\label{fig:nondelayall}
%\end{figure}

\begin{thebibliography}{99}
	
	
	\bibitem{alonscience} Milo, R., Shen-Orr, S., Itzkovitz, S., Kashtan, N., Chklovskii, D., $\&$ Alon, U. (2002). Network motifs: simple building blocks of complex networks. Science, 298(5594), 824-827.
	
	\bibitem{alon} Alon, U. (2007). Network motifs: theory and experimental approaches. Nature Reviews Genetics, 8(6), 450-461.
	
	\bibitem{mizuno}Mizuno-Yamasaki, E., Rivera-Molina, F.,$ \&$ Novick, P. (2012). GTPase networks in membrane traffic. Annual review of biochemistry, 81, 637-659.https://doi.org/10.1146/annurev-biochem-052810-093700
	
	\bibitem{barr} Barr, F. A. (2013). Rab GTPases and membrane identity: causal or inconsequential?. Journal of Cell Biology, 202(2), 191-199.
	
	
	\bibitem{colicelli} Colicelli, J. (2004). Human RAS superfamily proteins and related GTPases. Science's STKE, 2004(250), re13-re13.
	
	
	
	\bibitem{subramani} Subramani, D., $\&$  Alahari, S. K. (2010). Integrin-mediated function of Rab GTPases in cancer progression. Molecular cancer, 9(1), 1-9.
	
	\bibitem{hutagalung} Hutagalung AH, Novick PJ. 2011 Role of Rab GTPases in membrane traffic and cell physiology. Physiol. Rev. 91, 119–149. (doi:10.1152/physrev. 00059.2009)
	
	
	\bibitem{goody} Goody, R. S., Müller, M. P., $\&$ Wu, Y. W. (2017). Mechanisms of action of Rab proteins, key regulators of intracellular vesicular transport. Biological chemistry, 398(5-6), 565-575.
	
	\bibitem{cherfils} Cherfils, J., $\&$ Zeghouf, M. (2013). Regulation of small gtpases by gefs, gaps, and gdis. Physiological reviews, 93(1), 269-309.
	
	\bibitem{jiang} Jiang, R., Tu, Z., Chen, T., $\&$ Sun, F. (2006). Network motif identification in stochastic networks. Proceedings of the National Academy of Sciences, 103(25), 9404-9409.
	
	\bibitem{ehrmann} Ehrmann, A., Nguyen, B., $\&$ Seifert, U. (2019). Interlinked GTPase cascades provide a motif for both robust switches and oscillators. Journal of the Royal Society Interface, 16(157), 20190198.https://doi.org/10.1098/rsif.2019.0198 
	
	
	\bibitem{zhang} Zhang, D., $\&$ Ouyang, Q. (2021). Nonequilibrium Thermodynamics in Biochemical Systems and Its Application. Entropy, 23(3), 271.https://doi.org/10.3390/e23030271
	
	\bibitem{es} Esposito, M. (2020). Open questions on nonequilibrium thermodynamics of chemical reaction networks. Communications Chemistry, 3(1), 1-3.
	
	\bibitem{sto1} Seifert, U. (2008). Stochastic thermodynamics: principles and perspectives. The European Physical Journal B, 64(3), 423-431.
	
	\bibitem{sto2} Seifert, U. (2019). From stochastic thermodynamics to thermodynamic inference. Annual Review of Condensed Matter Physics, 10, 171-192.
	
	\bibitem{nicolis} Nicolis, G.; Prigogine, I. Self-Organization in Nonequilibrium Systems: From Dissipative Structures to Order through Fluctuations;
	Wiley: New York, NY, USA, 1977.
	
	\bibitem{prigogine} Prigogine, I. Introduction to Thermodynamics of Irreversible Processes; Wiley: New York, NY, USA, 1967.
	
	\bibitem{gingrich} Gingrich, T. R., Horowitz, J. M., Perunov, N., $\&$ England, J. L. (2016). Dissipation bounds all steady-state current fluctuations. Physical review letters, 116(12), 120601.
	
	\bibitem{horowitz} Horowitz, J. M., $\&$ Gingrich, T. R. (2020). Thermodynamic uncertainty relations constrain non-equilibrium fluctuations. Nature Physics, 16(1), 15-20.
	
	\bibitem{hasegawa} Hasegawa, Y., $\&$ Van Vu, T. (2019). Fluctuation theorem uncertainty relation. Physical review letters, 123(11), 110602.
	
	\bibitem{barato} Barato, A. C., $\&$ Seifert, U. (2015). Thermodynamic uncertainty relation for biomolecular processes. Physical review letters, 114(15), 158101. https://doi.org/10.1103/PhysRevLett.114.158101
	
	\bibitem{pietzonka} Pietzonka, P., Barato, A. C., $\&$ Seifert, U. (2016). Universal bound on the efficiency of molecular motors. Journal of Statistical Mechanics: Theory and Experiment, 2016(12), 124004.
	
	\bibitem{kolomeisky} Kolomeisky, A. B., $\&$ Fisher, M. E. (2007). Molecular motors: a theorist's perspective. Annu. Rev. Phys. Chem., 58, 675-695.
	
	\bibitem{bustamante} Bustamante, C., Keller, D., $\&$ Oster, G. (2001). The physics of molecular motors. Accounts of chemical research, 34(6), 412-420.
	
	\bibitem{cao} Cao, Y., Wang, H., Ouyang, Q., $\&$ Tu, Y. (2015). The free-energy cost of accurate biochemical oscillations. Nature physics, 11(9), 772-778.https://doi.org/10.1038/nphys3412
	
	\bibitem{marsland} Marsland III, R., Cui, W., $\&$ Horowitz, J. M. (2019). The thermodynamic uncertainty relation in biochemical oscillations. Journal of the Royal Society Interface, 16(154), 20190098.https://doi.org/10.1098/rsif.2019.0098
	
	\bibitem{wierenga} Wierenga, H., Ten Wolde, P. R., $\&$ Becker, N. B. (2018). Quantifying fluctuations in reversible enzymatic cycles and clocks. Physical Review E, 97(4), 042404.
	
	\bibitem{brownian}Barato, A. C., $\&$ Seifert, U. (2016). Cost and precision of Brownian clocks. Physical Review X, 6(4), 041053.
	
	\bibitem{copy} Song, Y., $\&$ Hyeon, C. (2020). Thermodynamic cost, speed, fluctuations, and error reduction of biological copy machines. The journal of physical chemistry letters, 11(8), 3136-3143. 
	
	\bibitem{bennett} Bennett, C. H. (1979). Dissipation-error tradeoff in proofreading. BioSystems, 11(2-3), 85-91.
	
	\bibitem{lan} Lan, G., Sartori, P., Neumann, S., Sourjik, V., $\&$ Tu, Y. (2012). The energy–speed–accuracy trade-off in sensory adaptation. Nature physics, 8(5), 422-428.
	
	
	
	\bibitem{glycolytic} Kim, P., $\&$ Hyeon, C. (2021). Thermodynamic optimality of glycolytic oscillations. The Journal of Physical Chemistry B.
	
	
	\bibitem{walczak} %Szymaska-Rożek, P., Villamaina, D., Miȩkisz, J., $\&$
	Walczak, A. M. (2019). Dissipation in non-steady state regulatory circuits. Entropy, 21(12), 1212.
	
	\bibitem{lee} Lee, S., Hyeon, C., $\&$ Jo, J. (2018). Thermodynamic uncertainty relation of interacting oscillators in synchrony. Physical Review E, 98(3), 032119.
	
	\bibitem{assess} Song, Y., $\&$ Hyeon, C. (2021). Thermodynamic uncertainty relation to assess biological processes. The Journal of Chemical Physics, 154(13), 130901.
	
	
	
	
	\bibitem{zhangg}Zhang, D., $\&$ Ouyang, Q. (2021). Nonequilibrium Thermodynamics in Biochemical Systems and Its Application. Entropy, 23(3), 271.
	
	
	\bibitem{koza} Koza, Z. (2000). Diffusion coefficient and drift velocity in periodic media. Physica A: Statistical Mechanics and its Applications, 285(1-2), 176-186. https://doi.org/10.1016/S0378-4371(00)00280-6
	
	\bibitem{koza1} Koza, Z. (1999). General technique of calculating the drift velocity and diffusion coefficient in arbitrary periodic systems. Journal of Physics A: Mathematical and General, 32(44), 7637.
	
	
	
	
	\bibitem{fanoo} Barato, A. C., $\&$ Seifert, U. (2015). Universal bound on the Fano factor in enzyme kinetics. The Journal of Physical Chemistry B, 119(22), 6555-6561. https://doi.org/10.1021/acs.jpcb.5b01918
	
	\bibitem{fastandslow} Zhang, X. P., Cheng, Z., Liu, F., $\&$ Wang, W. (2007). Linking fast and slow positive feedback loops creates an optimal bistable switch in cell signaling. Physical Review E, 76(3), 031924.
	
	\bibitem{chia} Chia, W. J., $\&$ Tang, B. L. (2009). Emerging roles for Rab family GTPases in human cancer. Biochimica et Biophysica Acta (BBA)-Reviews on Cancer, 1795(2), 110-116.
	
	\bibitem{maria} Ioannou, M. S., $\&$ McPherson, P. S. (2016). Regulation of cancer cell behavior by the small GTPase Rab13. Journal of Biological Chemistry, 291(19), 9929-9937.
	
	\bibitem{li} Li, G. (2011). Rab GTPases, membrane trafficking and diseases. Current drug targets, 12(8), 1188-1193.
	
	\bibitem{muller} Müller, M. P., $\&$ Goody, R. S. (2018). Molecular control of Rab activity by GEFs, GAPs and GDI. Small GTPases, 9(1-2), 5-21.
	
	\bibitem{stein} Stein, M. P., Dong, J., $\&$ Wandinger-Ness, A. (2003). Rab proteins and endocytic trafficking: potential targets for therapeutic intervention. Advanced drug delivery reviews, 55(11), 1421-1437.
	
	\bibitem{guadagno} Guadagno, N. A., $\&$ Progida, C. (2019). Rab GTPases: switching to human diseases. Cells, 8(8), 909.
	
	\bibitem{chen} Chen, Y., Ng, F., $\&$ Tang, B. L. (2016). Rab23 activities and human cancer—emerging connections and mechanisms. Tumor Biology, 37(10), 12959-12967.
	
	\bibitem{romano} Romano, G., Nigita, G., Calore, F., Saviana, M., Le, P., Croce, C. M., ... $\&$ Nana-Sinkam, P. (2020). MiR-124a Regulates Extracellular Vesicle Release by Targeting GTPase Rabs in Lung Cancer. Frontiers in Oncology, 10.
	
	\bibitem{yang} Yang, X. Z., Li, X. X., Zhang, Y. J., Rodriguez-Rodriguez, L., Xiang, M. Q., Wang, H. Y.,$\&$ Zheng, X. S. (2016). Rab1 in cell signaling, cancer and other diseases. Oncogene, 35(44), 5699-5704.
	
	\bibitem{tzeng} Tzeng, H. T., $\&$ Wang, Y. C. (2016). Rab-mediated vesicle trafficking in cancer. Journal of biomedical science, 23(1), 1-7.
	
	\bibitem{oral} Zhang, D., Lu, C., $\&$ Ai, H. (2017). Rab5a is overexpressed in oral cancer and promotes invasion through ERK/MMP signaling. Molecular medicine reports, 16(4), 4569-4576.
	
	\bibitem{xin}Qin, X., Wang, J., Wang, X., Liu, F., Jiang, B., $\&$ Zhang, Y. (2017). Targeting Rabs as a novel therapeutic strategy for cancer therapy. Drug discovery today, 22(8), 1139-1147.
	
	\bibitem{delprato} Delprato, A. (2012). Topological and functional properties of the small GTPases protein interaction network.
	
	\bibitem{ali} Ali, S., Malik, M. Z., Singh, S. S., Chirom, K., Ishrat, R., $\&$ Singh, R. B. (2018). Exploring novel key regulators in breast cancer network. PLoS One, 13(6), e0198525.
	
	\bibitem{rocky} Mangangcha, I. R., Malik, M. Z., Küçük, Ö., Ali, S., $\&$ Singh, R. B. (2019). Identification of key regulators in prostate cancer from gene expression datasets of patients. Scientific reports, 9(1), 1-16.
	
	\bibitem{malik} Malik, M. Z., Chirom, K., Ali, S., Ishrat, R., Somvanshi, P., $\&$ Singh, R. B. (2019). Methodology of predicting novel key regulators in ovarian cancer network: a network theoretical approach. BMC cancer, 19(1), 1-16.
	
	\bibitem{andre} Andreopoulos, B., Winter, C., Labudde, D., $\&$ Schroeder, M. (2009). Triangle network motifs predict complexes by complementing high-error interactomes with structural information. BMC bioinformatics, 10(1), 1-20.
	
	\bibitem{cloutier} Cloutier, M., $\&$ Wang, E. (2011). Dynamic modeling and analysis of cancer cellular network motifs. Integrative Biology, 3(7), 724-732.
	
	\bibitem{jeon} Jeon, H., Kim, S. R., Nam, D., $\&$ Yoo, Y. J. (2017). Analysis of triangular motifs in protein interaction networks and their implications to protein ages and cancer genes. International Journal of Data Mining and Bioinformatics, 19(4), 340-365.
	
	\bibitem{schramm} Schramm, G., Kannabiran, N., $\&$ König, R. (2010). Regulation patterns in signaling networks of cancer. BMC systems biology, 4(1), 1-12.
	
	\bibitem{krishnan}Gopal Krishnan, P. D., Golden, E., Woodward, E. A., Pavlos, N. J., $\&$ Blancafort, P. (2020). Rab GTPases: emerging oncogenes and tumor suppressive regulators for the editing of survival pathways in cancer. Cancers, 12(2), 259.
	
	
	
	\bibitem{jordens} Jordens, I., Marsman, M., Kuijl, C.,$\&$ Neefjes, J. (2005). Rab proteins, connecting transport and vesicle fusion. Traffic, 6(12), 1070-1077.
	
	
	\bibitem{qian} Qian, H., $\&$ Reluga, T. C. (2005). Nonequilibrium thermodynamics and nonlinear kinetics in a cellular signaling switch. Physical review letters, 94(2), 028101.
	
	\bibitem{kim1} Kim, J. R., Yoon, Y., $\&$ Cho, K. H. (2008). Coupled feedback loops form dynamic motifs of cellular networks. Biophysical journal, 94(2), 359-365.
	
	\bibitem{anan} Ananthasubramaniam, B., $\&$ Herzel, H. (2014). Positive feedback promotes oscillations in negative feedback loops. PLoS One, 9(8), e104761.
	
	
	\bibitem{klinke} Klinke, D. J., Horvath, N., Cuppett, V., Wu, Y., Deng, W., $\&$  Kanj, R. (2015). Interlocked positive and negative feedback network motifs regulate $\beta$-catenin activity in the adherens junction pathway. Molecular biology of the cell, 26(22), 4135-4148.
	
	\bibitem{kim} Kim, D., Kwon, Y. K., $\&$ Cho, K. H. (2007). Coupled positive and negative feedback circuits form an essential building block of cellular signaling pathways. BioEssays, 29(1), 85-90.
	
	
	\bibitem{ferrell} Ferrell Jr, J. E., $\&$ Ha, S. H. (2014). Ultrasensitivity part III: cascades, bistable switches, and oscillators. Trends in biochemical sciences, 39(12), 612-618.
	
	
	
	\bibitem{tian1} Tian, X. J., Zhang, X. P., Liu, F., $\&$ Wang, W. (2009). Interlinking positive and negative feedback loops creates a tunable motif in gene regulatory networks. Physical Review E, 80(1), 011926.
	
	
	
	\bibitem{tsai} Tsai, T. Y. C., Choi, Y. S., Ma, W., Pomerening, J. R., Tang, C., $\&$ Ferrell, J. E. (2008). Robust, tunable biological oscillations from interlinked positive and negative feedback loops. Science, 321(5885), 126-129.
	
	\bibitem{junco}Del Junco, C., $\&$ Vaikuntanathan, S. (2020). High chemical affinity increases the robustness of biochemical oscillations. Physical Review E, 101(1), 012410.
	
	\bibitem{turing} Turing, A. M. (1990). The chemical basis of morphogenesis. Bulletin of mathematical biology, 52(1), 153-197.
	
	\bibitem{kholodenko} Kholodenko, B. N. (2006). Cell-signalling dynamics in time and space. Nature reviews Molecular cell biology, 7(3), 165-176.
	
	\bibitem{graessl} Graessl, M., Koch, J., Calderon, A., Kamps, D., Banerjee, S., Mazel, T., ... $\&$ Nalbant, P. (2017). An excitable Rho GTPase signaling network generates dynamic subcellular contraction patterns. Journal of Cell Biology, 216(12), 4271-4285.
		\bibitem{yu} Yu, Q., Zhang, D., $\&$ Tu, Y. (2021). Inverse Power Law Scaling of Energy Dissipation Rate in Nonequilibrium Reaction Networks. Physical Review Letters, 126(8), 080601.
	
	\bibitem{rana} Rana, S., $\&$ Barato, A. C. (2020). Precision and dissipation of a stochastic turing pattern. Physical Review E, 102(3), 032135.
	
	\bibitem{denk} Denk, J., Kretschmer, S., Halatek, J., Hartl, C., Schwille, P., $\&$ Frey, E. (2018). MinE conformational switching confers robustness on self-organized Min protein patterns. Proceedings of the National Academy of Sciences, 115(18), 4553-4558.
	

	
\end{thebibliography}
\end{document}